\documentclass[twocolumn,eqsecnum,aps]{revtex4}
\usepackage[dvips]{graphicx}
\usepackage{amsmath,amsfonts,amsbsy,amssymb}

\input{tcilatex}

\begin{document}

\title{SU(4) Skyrmions and Activation Energy Anomaly \\
in Bilayer Quantum Hall Systems }
\author{Z.F. Ezawa$^1$ and G. Tsitsishvili$^2$}
\affiliation{{}$^1$Department of Physics, Tohoku University, Sendai, 980-8578 Japan }
\affiliation{{}$^2$Department of Theoretical Physics, A. Razmadze Mathematical Institute,
Tbilisi, 380093 Georgia}
\date{\today }

\begin{abstract}
The bilayer QH system has four energy levels in the lowest Landau level,
corresponding to the layer and spin degrees of freedom. We investigate the
system in the regime where all four levels are nearly degenerate and equally
active. The underlying group structure is SU(4). At $\nu =1$ the QH state is
a charge-transferable state between the two layers and the SU(4) isospin
coherence develops spontaneously. Quasiparticles are isospin textures to be
identified with SU(4) skyrmions. The skyrmion energy consists of the Coulomb
energy, the Zeeman energy and the pseudo-Zeeman energy. The Coulomb energy
consists of the self-energy, the capacitance energy and the exchange energy.
At the balanced point only pseudospins are excited unless the tunneling gap
is too large. Then, the SU(4) skyrmion evolves continuously from the
pseudospin-skyrmion limit into the spin-skyrmion limit as the system is
transformed from the balanced point to the monolayer point by controlling
the bias voltage. Our theoretical result explains quite well the
experimental data due to Murphy et al. and Sawada et al. on the activation
energy anomaly induced by applying parallel magnetic field.
\end{abstract}

\maketitle

\section{Introduction}

Exchange Coulomb interactions play key roles in various strongly correlated
electron systems. They are essential also in quantum Hall (QH) systems\cite%
{BookDasSarma,BookEzawa}, where spins are polarized spontaneously as in
ferromagnets. Much more interesting phenomena occur in bilayer QH systems
due to exchange interactions. For instance, an anomalous tunneling current
has been observed\cite{Spielman00L} between the two layers at the zero bias
voltage. It may well be a manifestation of the Josephson-like phenomena
predicted a decade ago\cite{Ezawa92IJMPB}. They occur due to quantum
coherence developed spontaneously across the layers\cite%
{Ezawa92IJMPB,Ezawa97B,Moon95B}.

Electrons in a plane perform cyclotron motion under strong magnetic field $%
B_{\perp }$ and create Landau levels. When one Landau level is filled up the
system becomes incompressible, leading to the integer QH effect. In QH
systems the electron position is specified solely by the guiding center $%
\mathbf{X}=(X,Y)$ subject to the noncommutative relation, $[X,Y]=-i\ell
_{B}^{2}$, with the magnetic length $\ell _{B}=\sqrt{\hbar /eB_{\perp }}$.
The QH system provides us with a simple realization of noncommutative
geometry. It follows from this relation that each electron occupies an area $%
2\pi \ell _{B}^{2}$ labelled by the Landau-site index. Electrons behave as
if they were on lattice sites, among which exchange interactions operate.

In QH ferromagnets charged excitations are spin textures called skyrmions%
\cite{Sondhi93B}. Skyrmions are identified experimentally\cite%
{Barrett95L,Aifer96L,Schmeller95L} by measuring the number of flipped spins
per one quasiparticle. Indeed, by tilting samples, the activation energy
increases by the Zeeman effect, which is roughly proportional to the number
of flipped spins. On the contrary, an entirely opposite behavior has been
observed by Murphy et al.\cite{Murphy94L} in the bilayer QH system at the
filling factor $\nu =1$, where the activation energy decreases rapidly by
tilting samples. It is called the activation energy anomaly because of this
unexpected behavior. Note that we expect an increase since the $\nu =1$
bilayer QH system is a QH ferromagnet with spins spontaneously polarized.
This anomalous decrease has been argued to occur due to the loss of the
exchange energy of a pseudospin texture based on the bimeron model\cite%
{Moon95B}, where the spin degree of freedom is frozen. Here, the layer
degree of freedom is referred to as the pseudospin: It is said to be up
(down) when an electron is in the front (back) layer.

We investigate physics taking place in the lowest Landau level (LLL). Since
each Landau site can accommodate four electrons with the spin and layer
degrees of freedom, the underlying group structure is enlarged to SU(4). Let
us call it the isospin SU(4) in contrast to the spin SU(2) and the
pseudospin SU(2). We study charged excitations in the $\nu =1$ bilayer QH
system. A natural candidate is the SU(4) isospin texture to be identified
with the SU(4) skyrmion\cite{Ezawa99L,BookEzawa}. A specific feature of the
SU(4) isospin texture is that it is reduced to the spin and pseudospin
textures in certain limits.

At $\nu =1$, electrons are transferable between the two layers continuously
without breaking the QH effect\cite{Sawada97SSC}. Namely, the bilayer QH
system can be continuously brought into the monolayer QH system by changing
the density imbalance. It implies that a pseudospin texture at the balanced
point must be continuously transformed into a spin texture at the monolayer
point. It is natural that a pseudospin texture evolves into an isospin
texture and then regresses to a spin texture in this process. The
corresponding continuous transformation of the activation energy has already
been observed experimentally by Sawada et al.\cite%
{Sawada97SSC,SawadaX03PE,Terasawa04}. In this paper we explain these
experimental data\cite{Murphy94L,SawadaX03PE,Terasawa04} based on the
excitation of SU(4) skyrmions.

In Section II we derive the Landau-site Hamiltonian governing bilayer QH
systems, by extending the algebraic method employed previously to
investigate multicomponent monolayer QH systems\cite{EzawaX03B}. In Section
III we rewrite the Landau-site Hamiltonian into the exchange-interaction
form. The Coulomb potentials associated with the direct and exchange
interactions are obtained analytically.

In Section IV the ground state is explored in the regime where the SU(4)
isospin coherence develops spontaneously. It is shown that the capacitance
energy consists of two terms arising from the direct interaction, which is
the standard one made of two planes, and from the exchange interaction, by
way of which the capacitance energy becomes considerably smaller than the
standard one for a small layer separation.

In Section V the excitation energy of an electron-hole pair is calculated
exactly. A hole (electron) is the small size limit of a skyrmion
(antiskyrmion), which is realized when the Zeeman gap $\Delta _{\text{Z}}$ \
and the tunneling gap $\Delta _{\text{SAS}}$ are very large. The exchange
energy is very different whether the pair excites the spin or the
pseudospin. Due to this difference the pseudospin excitation occurs at the
balanced point even if the tunneling gap is quite large. Our result explains
why pseudospin flips were observed in the experiment due to Terasawa et al.%
\cite{Terasawa04}\ in a sample having a very large tunneling gap ($\Delta _{%
\text{SAS}}\simeq 33$K). In Section VI we study skyrmions in a microscopic
theory, extending the approach\cite{Fertig94B} previously known for the
monolayer system.

In Section VII the effective Hamiltonian is derived from the Landau-site
Hamiltonian by making the derivative expansion. We introduce composite
bosons together with the CP$^{3}$ field\cite{Ezawa99L} to describe coherent
excitations. In Section VIII we carry out an analysis of SU(4) isospin
textures identified with topological excitations called CP$^{3}$ skyrmions
or equivalently SU(4) skyrmions. A spin texture (spin-skyrmion) and a
pseudospin texture (ppin-skyrmion) are special limits of an SU(4) skyrmion.
We estimate the excitation energy of one SU(4) skyrmion. We show that, if a
ppin-skyrmion is excited at the balanced point, it evolves continuously into
a spin-skyrmion at the monolayer point via an SU(4) skyrmion as the density
imbalance is increased between the two layers.

In Section IX we investigate the activation energy anomaly by applying the
parallel magnetic field between the two layers. We calculate explicitly the
loss of exchange energy of one SU(4) skyrmion, which is shown to be
proportional to the capacitance term. Our theoretical result explains quite
well both experimental data due to Murphy et al.\cite{Murphy94L} and Sawada
et al.\cite{SawadaX03PE,Terasawa04}.

Section X is devoted to discussions.

\section{Landau-Site Hamiltonian}

\label{SecLandaHamil}

Electrons perform cyclotron motion under perpendicular magnetic field $%
B_{\bot }$. The number of flux quanta passing through the system is $N_{\Phi
}\equiv B_{\perp }S/\Phi _{\text{D}}$, where $S$ is the area and $\Phi _{%
\text{D}}=2\pi \hbar /e$ is the flux quantum. There are $N_{\Phi }$ Landau
sites per one Landau level, each of which is associated with one flux
quantum and occupies an area $S/N_{\Phi }=2\pi \ell _{B}^{2}$. In the
bilayer system an electron has two types of indices, the layer index (f, b)
and the spin index ($\uparrow ,\downarrow $). One Landau site may
accommodate four electrons. The filling factor is $\nu =N/N_{\Phi }$ with $N$
the total number of electrons.

The kinetic Hamiltonian is%
\begin{equation}
H_{\text{K}}={\frac{1}{2M}}\int \!d^{2}x\;\Psi ^{\dag }(\mathbf{x}%
)(D_{x}-iD_{y})(D_{x}+iD_{y})\Psi (\mathbf{x}),  \label{HamilKinem}
\end{equation}%
where $\Psi $ stands for the four-component electron field, and $%
D_{k}=-i\hbar \partial _{k}+eA_{k}$ is the covariant momentum. The kinetic
Hamiltonian, which is invariant under the global SU(4) transformation,
creates Landau levels. Assuming a large Landau level separation we focus on
physics taking place in the LLL. Then we may neglect the kinetic energy
since it is common to all states.

We introduce $4\times 4$ matrices $\tau _{a}^{\text{spin}}$ and $\tau _{a}^{%
\text{ppin}}$ generating the spin and pseudospin (ppin) SU(2) groups, as
summarized in Appendix A; in particular, $\tau _{z}^{\text{spin}}=$diag$%
(1,-1,1,-1)$ and $\tau _{z}^{\text{ppin}}=$diag$(1,1,-1,-1)$. To avoid
confusions we use $S_{a}$ for the spin SU(2) field, $P_{a}$ for the
pseudospin SU(2) field ($a=x,y,z$) and $I_{A}$ for the isospin SU(4) field ($%
A=1,2,\cdots 15$).

The total Hamiltonian consists of the Coulomb term, the Zeeman term, the
tunneling term and the bias term. The role of the bias term is to transfer
electrons from one layer to the other by applying the bias voltage $V_{\text{%
bias}}$ between the two layers.

The Coulomb interaction is decomposed into the SU(4)-invariant and
SU(4)-noninvariant terms, 
\begin{align}
H_{\text{C}}^{+}& ={\frac{1}{2}}\int \!d^{2}xd^{2}y\;V^{+}(\mathbf{x}-%
\mathbf{y})\rho (\mathbf{x})\rho (\mathbf{y}),  \label{CouloSU4P} \\
H_{\text{C}}^{-}& ={2}\int \!d^{2}xd^{2}y\;V^{-}(\mathbf{x}-\mathbf{y})P_{z}(%
\mathbf{x})P_{z}(\mathbf{y}),  \label{CouloSU4M}
\end{align}%
where $\rho (\mathbf{x})=\Psi ^{\dag }(\mathbf{x})\Psi (\mathbf{x})$, $%
2P_{a}(\mathbf{x})=\Psi ^{\dag }(\mathbf{x})\tau _{a}^{\text{ppin}}\Psi (%
\mathbf{x})$ and 
\begin{equation}
V^{\pm }(\mathbf{x})={\frac{e^{2}}{8\pi \varepsilon }}\left( {\frac{1}{|%
\mathbf{x}|}\pm }\frac{1}{\sqrt{|\mathbf{x}|^{2}+d^{2}}}\right)
\label{CouloPotenBi}
\end{equation}%
with the layer separation $d$.

The Zeeman term is%
\begin{equation}
\mathcal{H}_{\text{Z}}=-\Delta _{\text{Z}}S_{z}(\mathbf{x),}
\label{ZeemaInter}
\end{equation}%
where $\Delta _{\text{Z}}$ is the Zeeman gap and $2S_{a}=\Psi ^{\dag }\tau
_{a}^{\text{spin}}\Psi $. The tunneling and bias terms are combined into the
pseudo-Zeeman term, 
\begin{equation}
\mathcal{H}_{\text{PZ}}=-\Delta _{\text{SAS}}P_{x}(\mathbf{x})-eV_{\text{bias%
}}P_{z}(\mathbf{x}),  \label{PseudZeemaInter}
\end{equation}%
where $\Delta _{\text{SAS}}$ is the tunneling gap.

The total Hamiltonian is%
\begin{equation}
H=H_{\text{C}}^{+}+H_{\text{C}}^{-}+H_{\text{Z}}+H_{\text{PZ}}.
\label{TotalHamil}
\end{equation}%
We investigate the regime where the SU(4)-invariant Coulomb term $H_{\text{C}%
}^{+}$ dominates all other interactions. Note that the SU(4)-noninvariant
Coulomb term vanishes, $H_{\text{C}}^{-}\rightarrow 0$, in the limit $%
d\rightarrow 0$.

We expand the electron field operator by a complete set of one-body wave
functions $\varphi _{i}(\mathbf{x})=\langle \mathbf{x}|i\rangle $ in the LLL,%
\begin{equation}
\psi _{\alpha }(\mathbf{x})\equiv \sum_{i=1}^{N_{\Phi }}c_{\alpha
}(i)\varphi _{i}(\mathbf{x}),  \label{ProjeField}
\end{equation}%
where $c_{\alpha }(i)$ is the annihilation operator at the Landau site $%
|i\rangle $ with $\alpha =$f$\uparrow $, f$\downarrow $, b$\uparrow $ and b$%
\downarrow $, 
\begin{align}
\{c_{\alpha }(i),c_{\beta }^{\dagger }(j)\}& =\delta _{ij}\delta _{\alpha
\beta },  \notag \\
\{c_{\alpha }(i),c_{\beta }(j)\}& =\{c_{\alpha }^{\dagger }(i),c_{\beta
}^{\dagger }(j)\}=0.  \label{AntiCommuC}
\end{align}%
It is impossible to choose an orthonormal complete set of one-body wave
functions $\varphi _{i}(\mathbf{x})$ in the LLL\cite{Iso92PLB}. Hence $\psi
_{\alpha }(\mathbf{x})$ does not satisfy the standard canonical
anticommutation relation, as implies that an electron cannot be localized to
a point within the LLL. Various interactions are projected to the LLL\cite%
{Girvin84B,Iso92PLB,Cappelli93NPB,EzawaX03B} by expanding the electron field
as in (\ref{ProjeField}).

The projected density is given by $\rho (\mathbf{x})=\Psi ^{\dag }(\mathbf{x}%
)\Psi (\mathbf{x})$ with the use of the field operator (\ref{ProjeField}).
Its Fourier transformation reads\cite{EzawaX03B}%
\begin{equation}
\rho (\mathbf{q})={\frac{1}{2\pi }}e^{-\ell _{B}^{2}\mathbf{q}%
^{2}/4}\sum_{mn}\langle m|e^{-i\mathbf{qX}}|n\rangle c_{m}^{\dag }c_{n},
\label{ProjeDensiQ}
\end{equation}%
where $\mathbf{X}=(X,Y)$ is the guiding center obeying the noncommutativity $%
[X,Y]=-i\ell _{B}^{2}$.\ Similar formulas hold for spin density operators
and so on. Substituting them into (\ref{CouloSU4P}) and (\ref{CouloSU4M}) we
find the projected Coulomb terms to be%
\begin{align}
H_{\text{C}}^{+}=\sum_{mnij}{V}_{mnij}^{+}& \sum_{\alpha \beta }c_{\alpha
}^{\dagger }(m)c_{\beta }^{\dagger }(i)c_{\beta }(j)c_{\alpha }(n),
\label{HamilPinC} \\
H_{\text{C}}^{-}=\sum_{mnij}{V}_{mnij}^{-}& \sum_{\alpha \beta \gamma \delta
}[\tau _{z}^{\text{ppin}}]_{\alpha \beta }[\tau _{z}^{\text{ppin}}]_{\gamma
\delta }  \notag \\
& \times c_{\alpha }^{\dagger }(m)c_{\gamma }^{\dagger }(i)c_{\delta
}(j)c_{\beta }(n)  \label{HamilMinC}
\end{align}%
with%
\begin{equation}
{V}_{mnij}^{\pm }={\frac{1}{4\pi }}\int d^{2}p\,V_{\text{D}}^{\pm }(\mathbf{p%
})\langle m|e^{i\mathbf{Xp}}|n\rangle \langle i|e^{-i\mathbf{Xp}}|j\rangle
\label{Vmnij}
\end{equation}%
and%
\begin{equation}
V_{\text{D}}^{\pm }(\mathbf{p})=\frac{e^{2}}{8\pi \varepsilon }\frac{1\pm
e^{-|\mathbf{p}|d}}{|\mathbf{p}|}e^{-\ell _{B}^{2}\mathbf{p}^{2}/2}.
\label{ProjeCoulo}
\end{equation}%
We also find%
\begin{align}
H_{\text{Z}}=& -\Delta _{\text{Z}}\sum_{m}S_{z}(m,m), \\
H_{\text{PZ}}=& -\sum_{m}\left[ \Delta _{\text{SAS}}P_{x}(m,m)+eV_{\text{bias%
}}P_{z}(m,m)\right] ,
\end{align}%
where%
\begin{align}
\rho (m,n)=& \sum_{\alpha }c_{\alpha }^{\dagger }(m)c_{\alpha }(n),  \notag
\\
S_{a}(m,n)=& \frac{1}{2}\sum_{\alpha \beta }c_{\alpha }^{\dagger }(m)[\tau
_{a}^{\text{spin}}]_{\alpha \beta }c_{\beta }(n),
\end{align}%
and a similar formula for $P_{a}(m,n)$. These formulas are derived precisely
in the same way as for the multicomponent monolayer system with the
replacement of the potential ${V}_{mnij}$ by ${V}_{mnij}^{+}$: See Section V
in Ref.\cite{EzawaX03B}.

For a later convenience we represent the projected Coulomb Hamiltonians (\ref%
{HamilPinC}) and (\ref{HamilMinC}) as%
\begin{align}
H_{\text{D}}^{+}=& \sum_{mnij}{V}_{mnij}^{+}\rho (m,n)\rho (i,j),
\label{HamilPInD} \\
H_{\text{D}}^{-}=& 4\sum_{mnij}{V}_{mnij}^{-}P_{z}(m,n)P_{z}(i,j).
\label{HamilMInD}
\end{align}%
In the momentum space the total Coulomb Hamiltonian reads%
\begin{align}
H_{\text{D}}=& \pi \int d^{2}p\,V_{\text{D}}^{+}(\mathbf{p})\widehat{\rho }(-%
\mathbf{p})\widehat{\rho }(\mathbf{p}),  \notag \\
& +4\pi \int {\!}d^{2}p{\,}V_{\text{D}}^{-}(\mathbf{p})\widehat{P}_{z}(-%
\mathbf{p})\widehat{P}_{z}(\mathbf{p}),  \label{DirecForm}
\end{align}%
where we have defined%
\begin{align}
\widehat{\rho }(\mathbf{p})=& \frac{1}{2\pi }\sum_{mn}^{\infty }\langle
m|e^{-i\mathbf{pX}}|n\rangle \rho (m,n),  \notag \\
\widehat{P}_{a}(\mathbf{p})=& \frac{1}{2\pi }\sum_{mn}^{\infty }\langle
m|e^{-i\mathbf{pX}}|n\rangle P_{a}(m,n).  \label{WeylOrderDensiX}
\end{align}%
We call (\ref{DirecForm}) the direct-interaction form of the Coulomb
Hamiltonian $\widehat{H}_{\text{C}}$.

Note the relation%
\begin{equation}
\rho (\mathbf{q})=e^{-\ell _{B}^{2}\mathbf{q}^{2}/4}\widehat{\rho }(\mathbf{q%
}).  \label{Naive2Weyl}
\end{equation}%
between the two types of densities (\ref{ProjeDensiQ}) and (\ref%
{WeylOrderDensiX}). Though $\widehat{\rho }(\mathbf{p})$ presents a useful
tool to work with noncommutative geometry\cite{EzawaX03B}, it turns out that 
$\rho (\mathbf{q})$ describes the physical density [see (\ref%
{Naive2WeylClass})].

The density operator $\widehat{\rho }(\mathbf{p})$ satisfies the W$_{\infty
} $ algebra\cite{Girvin86B,Iso92PLB,Cappelli93NPB}, 
\begin{equation}
\lbrack \widehat{\rho }(\mathbf{p}),\widehat{\rho }(\mathbf{q})]={\frac{i}{%
\pi }}\widehat{\rho }(\mathbf{p}+\mathbf{q})\sin \left[ \frac{1}{2}\ell
_{B}^{2}\mathbf{p}\!\wedge \!\mathbf{q}\right] .  \label{AlgebW}
\end{equation}%
For the SU(4) isospin field we may define%
\begin{equation}
I_{A}(m,n)=\frac{1}{2}\sum_{\alpha \beta }c_{\alpha }^{\dagger }(m)[\lambda
_{A}]_{\alpha \beta }c_{\beta }(n),  \label{DensiImn}
\end{equation}%
where $\lambda _{A}$ is the generating matrices [See Appendix A], and%
\begin{equation}
\widehat{I}_{A}(\mathbf{p})=\frac{1}{2\pi }\sum_{mn}^{\infty }\langle m|e^{-i%
\mathbf{pX}}|n\rangle I_{A}(m,n).  \label{DensiIp}
\end{equation}%
All of the density operators $\widehat{\rho }(\mathbf{p})$ and $\widehat{I}%
_{A}(\mathbf{p})$\ form an algebra, which we have called\cite{EzawaX03B} W$%
_{\infty }$(4) since it is the SU(4) extension of W$_{\infty }$. It gives
the fundamental algebraic structure of the noncommutative system made of
4-component electrons. In general there arises the W$_{\infty }$(N) algebra
in the N-component QH ferromagnet.

We define classical densities $\widehat{\rho }^{\text{cl}}(\mathbf{p})$, $%
\widehat{I}_{A}^{\text{cl}}(\mathbf{p})$, $\widehat{S}_{a}^{\text{cl}}(%
\mathbf{p})$ and $\widehat{P}_{a}^{\text{cl}}(\mathbf{p})$ by%
\begin{equation}
\widehat{\rho }^{\text{cl}}(\mathbf{p})=\langle \mathfrak{S}|\widehat{\rho }(%
\mathbf{p})|\mathfrak{S}\rangle ,\quad \widehat{I}_{A}^{\text{cl}}=\langle 
\mathfrak{S}|\widehat{I}_{A}(\mathbf{p})|\mathfrak{S}\rangle ,
\label{ClassDensi}
\end{equation}%
and so on, where $|\mathfrak{S}\rangle $ represents a skyrmion state. In the
coordinate space the relation%
\begin{equation}
\sum_{A=1}^{N^{2}-1}\widehat{I}_{A}^{\text{cl}}(\mathbf{x})\bigstar \widehat{%
I}_{A}^{\text{cl}}(\mathbf{x})=\frac{1}{2N}\widehat{\rho }^{\text{cl}}(%
\mathbf{x})\bigstar \left[ \frac{N}{2\pi \ell _{B}^{2}}-\widehat{\rho }^{%
\text{cl}}(\mathbf{x})\right]  \label{CasimInX}
\end{equation}%
holds among the classical densities associated with the generators of the W$%
_{\infty }$(N) algebra, where $\bigstar $\ stands for the Moyal star
product. We demonstrate this formula in Appendix \ref{AppenDandE}. Remark
the invariance of (\ref{CasimInX}) under 
\begin{equation}
\hat{\rho}^{\text{cl}}(\mathbf{r})\rightarrow \frac{N}{2\pi \ell _{B}^{2}}-%
\hat{\rho}^{\text{cl}}(\mathbf{r}).
\end{equation}%
This represents the electron-hole symmetry.

\section{Exchange Interactions}

In classical theory the Coulomb energy is simply given by (\ref{CouloSU4P})
with the use of the classical density $\rho ^{\text{cl}}(\mathbf{x})$, but
this is not the case in quantum theory. The exchange interaction emerges as
an important interaction from the exchange integral over wave functions. We
present a rigorous treatment of the direct and exchange energies valid in QH
systems.

For this purpose we rewrite the microscopic Coulomb Hamiltonians (\ref%
{HamilPinC}) and (\ref{HamilMinC}) into entirely different forms. Based on
an algebraic relation, 
\begin{equation}
\delta _{\sigma \beta }\delta _{\tau \alpha }=\frac{1}{2}\sum_{A}^{N^{2}-1}[%
\lambda _{A}]_{\sigma \tau }[\lambda _{A}]_{\alpha \beta }+\frac{1}{N}\delta
_{\sigma \tau }\delta _{\alpha \beta },  \label{AlgebDX}
\end{equation}%
which holds for SU(N) with $\lambda _{A}$ the generating matrices, these
Coulomb Hamiltonians are equivalent to

\begin{align}
H_{\text{X}}^{+}=& -2\sum_{mnij}{V}_{mnij}^{+}\big[%
\sum_{A=1}^{15}I_{A}(m,j)I_{A}(i,n)  \notag \\
& \hspace{0.8in}+\frac{1}{2N}\rho (m,j)\rho (i,n)\big],  \label{SU4InvarMN}
\end{align}%
and%
\begin{align}
H_{\text{X}}^{-}=& -2\sum_{mnij}V_{mnij}^{-}\big[\sum_{A=1}^{15}\xi
_{A}I_{A}(m,j)I_{A}(i,n)  \notag \\
& \hspace{0.8in}+\frac{1}{2N}\rho (m,j)\rho (i,n)\big],  \label{SU4NoninMN}
\end{align}%
where $N=4$, $I_{A}(m,n)$ is defined by (\ref{DensiImn}) and $\xi
_{1,2,3,8,13,14,15}=+1$, $\xi _{4,5,6,7,9,10,11,12}=-1$. In the momentum
space they read 
\begin{align}
H_{\text{X}}^{+}=& \pi \int d^{2}p\,V_{\text{X}}^{+}(\mathbf{p})\big[%
\sum_{A=1}^{15}\widehat{I}_{A}(-\mathbf{p})\widehat{I}_{A}(\mathbf{p}) 
\notag \\
& \hspace{1in}+\frac{1}{2N}\widehat{\rho }(-\mathbf{p})\widehat{\rho }(%
\mathbf{p})\big], \\
H_{\text{X}}^{-}=& 4\pi \int {\!}d^{2}p{\,}V_{\text{X}}^{-}(\mathbf{p})\big[%
\sum_{A=1}^{15}\xi _{A}\widehat{I}_{A}(-\mathbf{p})\widehat{I}_{A}(\mathbf{p}%
)  \notag \\
& \hspace{1in}+\frac{1}{2N}\widehat{\rho }(-\mathbf{p})\widehat{\rho }(%
\mathbf{p})\big],  \label{ExchaHamilX}
\end{align}%
where 
\begin{equation}
V_{\text{X}}^{\pm }(\mathbf{p})=\frac{\ell _{B}^{2}}{\pi }\int
\!d^{2}k\,e^{-i\ell _{B}^{2}\mathbf{p}\wedge \mathbf{k}}V_{\text{D}}^{\pm }(%
\mathbf{k})  \label{KerneX}
\end{equation}%
and $N=4$.

We change the SU(4) basis from $\lambda _{A}$ to $\tau _{a}^{\text{spin}}$, $%
\tau _{a}^{\text{ppin}}$ and $\tau _{a}^{\text{spin}}\tau _{b}^{\text{ppin}}$
by way of the formula (\ref{AppBasisSU4}) in Appendix, which transforms
variable $I_{A}(m,n)$ to a set of variables,%
\begin{align}
S_{a}(m,n)& =\frac{1}{2}\sum_{\sigma \tau }c_{\sigma }^{\dagger }(m)(\tau
_{a}^{\text{spin}})_{\sigma \tau }c_{\tau }(n),  \notag \\
P_{a}(m,n)& =\frac{1}{2}\sum_{\sigma \tau }c_{\sigma }^{\dagger }(m)(\tau
_{a}^{\text{ppin}})_{\sigma \tau }c_{\tau }(n),  \notag \\
R_{ab}(m,n)& =\frac{1}{2}\sum_{\sigma \tau }c_{\sigma }^{\dagger }(m)(\tau
_{a}^{\text{spin}}\tau _{b}^{\text{ppin}})_{\sigma \tau }c_{\tau }(n).
\label{SPRtoI}
\end{align}%
The sum of (\ref{SU4InvarMN}) and (\ref{SU4NoninMN}) is expressed as%
\begin{align}
H_{\text{X}}=& -\sum_{mnij}{V}%
_{mnij}^{d}[S_{a}(m,j)S_{a}(i,n)+P_{a}(m,j)P_{a}(i,n)  \notag \\
& \hspace{0.8in}+R_{ab}(m,j)R_{ab}(i,n)]  \notag \\
& -2\sum_{mnij}V_{mnij}^{-}[S_{a}(m,j)S_{a}(i,n)+P_{z}(m,j)P_{z}(i,n)  \notag
\\
& \hspace{0.8in}+R_{az}(m,j)R_{az}(i,n)]  \notag \\
& -\frac{1}{4}\sum_{mnij}{V}_{mnij}\rho (m,j)\rho (i,n),  \label{HamilSPR}
\end{align}%
where%
\begin{equation}
{V}_{mnij}={V}_{mnij}^{+}+{V}_{mnij}^{-},\quad {V}_{mnij}^{d}={V}_{mnij}^{+}-%
{V}_{mnij}^{-}.
\end{equation}%
In the momentum space we find 
\begin{align}
H_{\text{X}}=& -\frac{\pi }{2}\int {\!}d^{2}p\,V_{\text{X}}^{d}(\mathbf{p})[%
\widehat{S}_{a}(-\mathbf{p})\widehat{S}_{a}(\mathbf{p})+\widehat{P}_{a}(-%
\mathbf{p})\widehat{P}_{a}(\mathbf{p})  \notag \\
& \hspace{0.8in}+\widehat{R}_{ab}(-\mathbf{p})\widehat{R}_{ab}(\mathbf{p})] 
\notag \\
& -\pi \int {\!}d^{2}p\,V_{\text{X}}^{-}(\mathbf{p})[\widehat{S}_{a}(-%
\mathbf{p})\widehat{S}_{a}(\mathbf{p})+\widehat{P}_{z}(-\mathbf{p})\widehat{P%
}_{z}(\mathbf{p})  \notag \\
& \hspace{0.8in}+\widehat{R}_{az}(-\mathbf{p})\widehat{R}_{az}(\mathbf{p})] 
\notag \\
& -\frac{\pi }{8}\int {\!}d^{2}p{\,}V_{\text{X}}(\mathbf{p})\widehat{\rho }(-%
\mathbf{p})\widehat{\rho }(\mathbf{p}),  \label{TruncX}
\end{align}%
where%
\begin{equation}
V_{\text{X}}(\mathbf{p})=V_{\text{X}}^{+}(\mathbf{p})+V_{\text{X}}^{-}(%
\mathbf{p}),\quad V_{\text{X}}^{d}(\mathbf{p})=V_{\text{X}}^{+}(\mathbf{p}%
)-V_{\text{X}}^{-}(\mathbf{p}).
\end{equation}%
In (\ref{HamilSPR}) and (\ref{TruncX}) the summation ($\sum_{a=xyz}$,$%
\sum_{b=xyz}$) over the repeated indices $a$ and $b$ is understood. We call (%
\ref{TruncX}) the exchange-interaction form of the Coulomb Hamiltonian $%
\widehat{H}_{\text{C}}$.

We have demonstrated that the Landau-site Hamiltonian $H_{\text{C}}$
possesses two entirely different forms, the direct-interaction form $H_{%
\text{D}}$ given by (\ref{DirecForm}) and the exchange-interaction form $H_{%
\text{X}}$ given by (\ref{TruncX}). They are equivalent, $H_{\text{D}}=H_{%
\text{X}}$, as the microscopic Hamiltonian. In this paper we are interested
in the excitation energy $\langle \mathfrak{S}|H|\mathfrak{S}\rangle $ of a
skyrmion state $|\mathfrak{S}\rangle $. In a previous paper\cite{EzawaX03B}
we have presented a heuristic argument showing that 
\begin{equation}
\langle \mathfrak{S}|H_{\text{C}}|\mathfrak{S}\rangle =H_{\text{D}}^{\text{cl%
}}+H_{\text{X}}^{\text{cl}}.  \label{CintoDX}
\end{equation}%
Here $H_{\text{D}}^{\text{cl}}$ and $H_{\text{X}}^{\text{cl}}$ are the
Hamiltonians of the direct-interaction form (\ref{DirecForm}) and of the
exchange-interaction form (\ref{TruncX}), where the density operators $%
\widehat{\rho }(\mathbf{p})$, $\widehat{I}_{A}(\mathbf{p})$, $\cdots $ are
replaced by the classical ones $\widehat{\rho }^{\text{cl}}(\mathbf{p})$, $%
\widehat{I}_{A}^{\text{cl}}(\mathbf{p})$, $\cdots $. We present a proof of
this formula for the skyrmion state in Appendix \ref{AppenDandE}. Thus, the
energy of a charge excitation consists of two well-separated pieces, the
direct energy and the exchange energy.

\section{Ground State}

We first determine the ground state of the bilayer system in this section.
For simplicity we start with the spin-frozen SU(2) bilayer system. We are
concerned about the regime dominated by the SU(2)-invariant Coulomb
interaction $H_{\text{C}}^{+}$. Hence we define the ground state as an
eigenstate of $H_{\text{C}}^{+}$ given by (\ref{HamilPinC}), and treat all
other interactions as small perturbations.

The unperturbed ground-state energy is given by%
\begin{equation}
H_{\text{C}}^{+}|\text{g}\rangle =\sum_{jn}V_{jnnj}^{+}|\text{g}\rangle
=-\epsilon _{\text{X}}^{+}N_{\Phi }|\text{g}\rangle  \label{SiteGrounEnergP}
\end{equation}%
with%
\begin{align}
\epsilon _{\text{X}}^{\pm }=& \frac{1}{N_{\Phi }}\sum_{jn}V_{jnnj}^{\pm }={%
\frac{1}{4\pi }}\int d^{2}k\,V_{\text{D}}^{\pm }(\mathbf{k})  \notag \\
=& \frac{1}{2}\left[ 1\pm e^{d^{2}/2\ell _{B}^{2}}\text{erfc}\left( d/\sqrt{2%
}\ell _{B}\right) \right] \Delta _{\text{C}}^{0},  \label{ParamXPM}
\end{align}%
where%
\begin{equation}
\Delta _{\text{C}}^{0}=\frac{1}{2}\sqrt{\frac{\pi }{2}}\frac{e^{2}}{4\pi
\varepsilon \ell _{B}}.
\end{equation}%
We take $\Delta _{\text{C}}^{0}$ as the Coulomb energy unit in this paper.
The unperturbed system is equivalent to the monolayer SU(2) QH ferromagnet,
where all pseudospins are spontaneously polarized into one arbitrary
direction in the SU(2) pseudospin space. The ground state may be expressed as%
\begin{align}
|\text{g}\rangle =\prod_{n}& \left\{ e^{i\vartheta _{0}/2}\sqrt{{\frac{%
1+\sigma _{0}}{2}}}c_{\text{f}}^{\dagger }(n)\right.  \notag \\
& \left. +e^{-i\vartheta _{0}/2}\sqrt{{\frac{1-\sigma _{0}}{2}}}c_{\text{b}%
}^{\dagger }(n)\right\} |0\rangle ,  \label{SiteGrounNoBreak}
\end{align}%
by introducing two constant parameters $\sigma _{0}$ and $\vartheta _{0}$.
We call $\sigma _{0}$ the imbalance parameter since it represents the
density imbalance between the two layers. The ground state is a coherent
state due to an infinite degeneracy with respect to $\sigma _{0}$ and $%
\vartheta _{0}$.

We then study the effect due to the SU(2)-noninvariant interaction. First,
due to the capacitance effect ($\varpropto V_{mnij}^{-}$) all pseudospins
are polarized into one arbitrary direction within the pseudospin $xy$ plane,
implying that the electron density is balanced between the two layers ($%
\sigma _{0}=0$). The symmetry SU(2) is broken into U(1). It is thus said
that the bilayer QH system is an easy plane ferromagnet. Next, we apply the
bias voltage to generate an imbalanced density state ($\sigma _{0}\neq 0$),
where the symmetry is still U(1). Finally, the tunneling interaction breaks
the symmetry completely by fixing $\vartheta _{0}=0$ in the state (\ref%
{SiteGrounNoBreak}). Consequently, the ground state is the bonding state $|$B%
$\rangle $ parametrized by the imbalance parameter $\sigma _{0}$, which is (%
\ref{SiteGrounNoBreak}) with $\vartheta _{0}=0$.

The generalization to the SU(4) bilayer system is trivial by adding the spin
component. The ground state turns out to be the up-spin bonding state $|$B$%
\uparrow \rangle $ due to the Zeeman effect. It is convenient to represent
the up-spin bonding state $|$B$\uparrow \rangle $ and the up-spin
antibonding state $|$A$\uparrow \rangle $\ as%
\begin{equation}
|\text{B\negthinspace }\uparrow \rangle =\prod\limits_{n}B_{\uparrow
}^{\dagger }(n)|0\rangle ,\quad |\text{A\negthinspace }\uparrow \rangle
=\prod\limits_{n}A_{\uparrow }^{\dagger }(n)|0\rangle ,
\end{equation}%
where%
\begin{equation}
\left( 
\begin{array}{c}
B_{\uparrow }(n) \\ 
A_{\uparrow }(n)%
\end{array}%
\right) ={\frac{1}{\sqrt{2}}}\left( 
\begin{array}{@{\,}rr}
\sqrt{1+\sigma _{0}} & \sqrt{1-\sigma _{0}} \\ 
\sqrt{1-\sigma _{0}} & -\sqrt{1+\sigma _{0}}%
\end{array}%
\right) \left( 
\begin{array}{c}
c_{\text{f}\uparrow }(n) \\ 
c_{\text{b}\uparrow }(n)%
\end{array}%
\right) .  \label{SiteOperaBAB}
\end{equation}%
The down-spin bonding state $|$B$\downarrow \rangle $ and the down-spin
antibonding state $|$A$\downarrow \rangle $\ are similarly defined.

We evaluate the ground state energy $E_{\text{g}}=\langle B\!\!\uparrow
|H|B\!\!\uparrow \rangle $ by representing the Hamiltonians in terms of
these operator. The ground-state energy per one site reads%
\begin{align}
E_{\text{g}}/N_{\Phi }& =-\epsilon _{\text{X}}^{+}+\frac{1}{4}\epsilon _{%
\text{cap}}\sigma _{0}^{2}  \notag \\
& -\frac{1}{2}\Delta _{\text{SAS}}\sqrt{1-\sigma _{0}^{2}}-\frac{1}{2}eV_{%
\text{bias}}\sigma _{0}-\Delta _{\text{Z}},  \label{GrounEnergSU4}
\end{align}%
where 
\begin{equation}
\epsilon _{\text{cap}}=4(\epsilon _{\text{D}}^{-}-\epsilon _{\text{X}}^{-})
\label{CapacParamDX}
\end{equation}%
with 
\begin{equation}
\epsilon _{\text{D}}^{-}=\frac{1}{N_{\Phi }}\sum_{jn}{V}_{nnjj}^{-}=\sqrt{%
\frac{1}{2\pi }}\frac{d}{\ell _{B}}\Delta _{\text{C}}^{0},
\end{equation}%
and $\epsilon _{\text{X}}^{-}$ given by (\ref{ParamXPM}). The imbalance
parameter $\sigma _{0}$ is determined to minimize the ground-state energy,%
\begin{equation}
\epsilon _{\text{cap}}\sigma _{0}+{\frac{\Delta _{\text{SAS}}\sigma _{0}}{%
\sqrt{1-\sigma _{0}^{2}}}}=eV_{\text{bias}},  \label{CondiOnSigma}
\end{equation}%
as a function of the bias voltage.

The second term in the ground-state energy (\ref{GrounEnergSU4}) is the
capacitance energy,%
\begin{equation}
E_{\text{g}}^{\text{cap}}=\frac{\epsilon _{\text{cap}}\sigma _{0}^{2}}{4}%
N_{\Phi }\equiv {\frac{Q^{2}}{2C}S,}  \label{CapacEnergX}
\end{equation}%
where $Q={\frac{1}{2}e}\rho _{0}\sigma _{0}$ is the charge imbalance and $C$
is the capacitance per unit area. We rewrite (\ref{CondiOnSigma}) as%
\begin{equation}
{\frac{Q}{C}}+V_{\text{junc}}=V_{\text{bias}},  \label{BiasSigmaX}
\end{equation}%
with%
\begin{equation}
eV_{\text{junc}}={\frac{\sigma _{0}}{\sqrt{1-\sigma _{0}^{2}}}}\Delta _{%
\text{SAS}}.  \label{BiasSASx}
\end{equation}%
When the tunneling interaction is absent ($\Delta _{\text{SAS}}\rightarrow 0$%
), (\ref{BiasSigmaX}) is reduced to 
\begin{equation}
Q=CV_{\text{bias}}.  \label{BiasVolta}
\end{equation}%
Eqs. (\ref{CapacEnergX}) and (\ref{BiasVolta}) are the well-known formulas
for the condenser, where the bias voltage $V_{\text{bias}}$ is balanced with
the electric potential $Q/C$ due to the charge difference $Q$ between the
two layers. In this case, a charge transfer between the two layers makes no
work, since there exists no potential difference between the two layers.

When $\Delta _{\text{SAS}}\neq 0$, the cancellation is imperfect as in (\ref%
{BiasSigmaX}). Thus, when a charge is moved from one layer to the other, it
is necessary to supply an energy against the potential difference $V_{\text{%
junc}}=V_{\text{bias}}-Q/C$. Consequently, the pseudo-Zeeman energy is given
by 
\begin{align}
\mathcal{H}_{\text{PZ}}& =-\left[ \Delta _{\text{SAS}}P_{x}(\mathbf{x})+eV_{%
\text{junc}}P_{z}(\mathbf{x})\right]  \notag \\
& =-\left[ P_{x}(\mathbf{x})+\frac{\sigma _{0}}{\sqrt{1-\sigma _{0}^{2}}}%
P_{z}(\mathbf{x})\right] \Delta _{\text{SAS}},  \label{PseudZeemaX}
\end{align}%
or%
\begin{equation}
H_{\text{PZ}}=-\sum_{m}\left[ P_{x}(m,m)+\frac{\sigma _{0}}{\sqrt{1-\sigma
_{0}^{2}}}P_{z}(m,m)\right] \Delta _{\text{SAS}}.  \label{PseudZeema}
\end{equation}%
As $\Delta _{\text{SAS}}\rightarrow 0$, the pseudo-Zeeman energy vanishes
even in imbalanced configuration ($\sigma _{0}\neq 0$), and the total
ground-state energy consists solely of the capacitance energy (\ref%
{CapacEnergX}). In order to analyze charge excitations it is necessary to
use (\ref{PseudZeema}) as the pseudo-Zeeman term in the total Hamiltonian (%
\ref{TotalHamil}).

\begin{figure}[t]
\begin{center}
\includegraphics[width=0.46\textwidth]{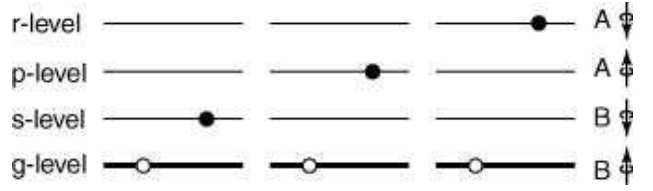}
\end{center}
\caption{ The LLL contains four energy levels corresponding to the two
layers and the two spin states. (We call them the g-level, s-level, p-level
and r-level.) At $\protect\nu =1$ the ground state is the up-spin bonding
state. An electron may be moved to any one of the other three levels to form
an electron-hole pair excitation.}
\label{FigLevelEH}
\end{figure}

\section{Electron and Hole Excitations}

\label{SecExcitEH}

As we see in the succeeding section, a skyrmion and an antiskyrmion are
reduced to a hole and an electron in their small size limit. We analyze
electron and hole excitations in this section to derive some exact results.

\subsection{Electron-Hole Pair Excitation}

One electron may be excited from the up-spin bonding state into the
down-spin bonding state, the up-spin antibonding state or the down-spin
antibonding state [FIG.\ref{FigLevelEH}]. These electron-hole states are%
\begin{align}
& |\psi _{\text{s}}\rangle =|\text{e}_{j}^{\text{B}\downarrow };\text{h}%
_{k}^{\text{B}\uparrow }\rangle =B_{\downarrow }^{\dag }(j)B_{\uparrow }(k)|%
\text{B\negthinspace \negthinspace }\uparrow \rangle , \\
& |\psi _{\text{p}}\rangle =|\text{e}_{j}^{\text{A}\uparrow };\text{h}_{k}^{%
\text{B}\uparrow }\rangle =A_{\uparrow }^{\dag }(j)B_{\uparrow }(k)|\text{%
B\negthinspace \negthinspace }\uparrow \rangle , \\
& |\psi _{\text{r}}\rangle =|\text{e}_{j}^{\text{A}\downarrow };\text{h}%
_{k}^{\text{B}\uparrow }\rangle =A_{\downarrow }^{\dag }(j)B_{\uparrow }(k)|%
\text{B\negthinspace \negthinspace }\uparrow \rangle .  \label{ThreeEHstate}
\end{align}%
We assume that an electron and a hole are separated far enough so that
interactions between them are neglected. The energy matrix looks as%
\begin{equation}
\left( 
\begin{array}{ccc}
\langle \psi _{\text{s}}|H|\psi _{\text{s}}\rangle & \langle \psi _{\text{s}%
}|H|\psi _{\text{r}}\rangle & 0 \\ 
\langle \psi _{\text{r}}|H|\psi _{\text{s}}\rangle & \langle \psi _{\text{r}%
}|H|\psi _{\text{r}}\rangle & 0 \\ 
0 & 0 & \langle \psi _{\text{p}}|H|\psi _{\text{p}}\rangle%
\end{array}%
\right) ,  \label{MatriSPR}
\end{equation}%
where%
\begin{align}
\langle \psi _{\text{s}}|H|\psi _{\text{s}}\rangle =& E_{\text{g}}+2\epsilon
_{\text{X}}^{+}+2\sigma _{0}^{2}\epsilon _{\text{X}}^{-}+\Delta _{\text{Z}},
\notag \\
\langle \psi _{\text{r}}|H|\psi _{\text{r}}\rangle =& E_{\text{g}}+2\epsilon
_{\text{X}}^{+}-2\sigma _{0}^{2}\epsilon _{\text{X}}^{-}-\sigma
_{0}^{2}\epsilon _{\text{cap}}+\Delta _{\text{Z}}+\frac{\Delta _{\text{SAS}}%
}{\sqrt{1-\sigma _{0}^{2}}},  \notag \\
\langle \psi _{\text{p}}|H|\psi _{\text{p}}\rangle =& E_{\text{g}}+2\epsilon
_{\text{X}}^{+}-2\epsilon _{\text{X}}^{-}-\sigma _{0}^{2}\epsilon _{\text{cap%
}}+\frac{\Delta _{\text{SAS}}}{\sqrt{1-\sigma _{0}^{2}}},  \notag \\
\langle \psi _{\text{r}}|H|\psi _{\text{s}}\rangle =& 2\sigma _{0}\sqrt{%
1-\sigma _{0}^{2}}\epsilon _{\text{D}}^{-}.  \label{PairEnergEH}
\end{align}%
We have $\langle \psi _{\text{s}}|H|\psi _{\text{p}}\rangle =\langle \psi _{%
\text{r}}|H|\psi _{\text{p}}\rangle =0$, because the Hamiltonian does not
flip the spin (it does not involve $S_{x}$ and $S_{y}$).

The matrix (\ref{MatriSPR}) is diagonal at the balanced point. The minimum
eigenvalue is the spin-excitation energy%
\begin{equation}
\langle \psi _{\text{s}}|H|\psi _{\text{s}}\rangle =E_{\text{g}}+2\epsilon _{%
\text{X}}^{+}+\Delta _{\text{Z}},  \label{PairEHs}
\end{equation}%
or the ppin excitation energy%
\begin{equation}
\langle \psi _{\text{p}}|H|\psi _{\text{p}}\rangle =E_{\text{g}}+2\epsilon _{%
\text{X}}^{+}-2\epsilon _{\text{X}}^{-}+\Delta _{\text{SAS}}.
\end{equation}%
The pseudospin excitation occurs when 
\begin{equation}
2\epsilon _{\text{X}}^{-}>\Delta _{\text{SAS}}-\Delta _{\text{Z}}.
\label{CondiPpinExcitEH}
\end{equation}%
We remark that $2\epsilon _{\text{X}}^{-}\simeq 47$K in a typical sample
with $d\simeq 231$ nm.

The matrix (\ref{MatriSPR}) is diagonal also at the monolayer point, where
the spin excitation always occurs with the excitation energy%
\begin{equation}
\langle \psi _{\text{s}}|H|\psi _{\text{s}}\rangle =E_{\text{g}}+2\left(
\epsilon _{\text{X}}^{+}+\epsilon _{\text{X}}^{-}\right) +\Delta _{\text{Z}},
\end{equation}%
since the energies of the other two modes diverge.

\begin{figure}[t]
\begin{center}
\includegraphics[width=0.46\textwidth]{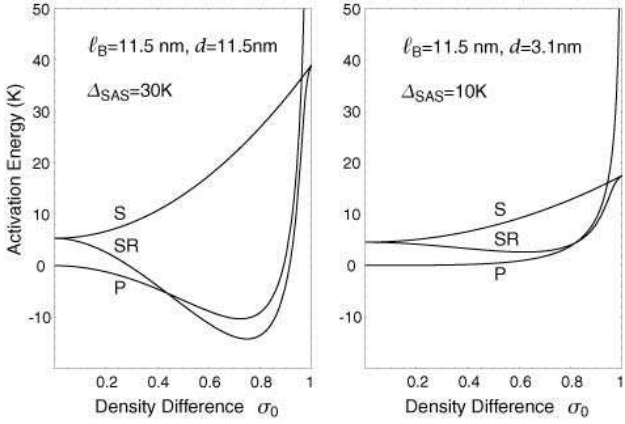}
\end{center}
\caption{The energy of an electron-hole pair state is depicted as a function
of the imbalance parameter $\protect\sigma_0$ based on the formulas ( 
\protect\ref{PairEnergEH}) and (\protect\ref{PairEnergSR}) with typical
sample parameters as indicated. Three curves corresponds to the state $|%
\protect\psi _{\text{s}}\rangle $, $|\protect\psi _{\text{sr}}\rangle $ and $%
|\protect\psi _{\text{p}}\rangle $. We have normalized the activation energy
of the pseudospin excitation to zero at $\protect\sigma _{0}=0$.}
\label{FigExcitEH}
\end{figure}

The two excitation states $|\psi _{\text{s}}\rangle $ and $|\psi _{\text{r}%
}\rangle $ mix to make a new state, $|\psi _{\text{sr}}\rangle $, to lower
the excitation energy except for $\sigma _{0}=0$ and $\sigma _{0}=1$. After
diagonalization it is%
\begin{align}
\langle \psi _{\text{sr}}|& H|\psi _{\text{sr}}\rangle =E_{\text{g}%
}+2\epsilon _{\text{X}}^{+}-\frac{1}{2}\sigma _{0}^{2}\epsilon _{\text{cap}%
}+\Delta _{\text{Z}}+\frac{1}{2}\widetilde{\Delta }_{\text{SAS}}  \notag \\
& -\frac{1}{2}\sqrt{(\widetilde{\Delta }_{\text{SAS}}-4\sigma
_{0}^{2}\epsilon _{\text{D}}^{-})^{2}+16\sigma _{0}^{2}(1-\sigma
_{0}^{2})(\epsilon _{\text{D}}^{-})^{2}}  \label{PairEnergSR}
\end{align}%
with $\widetilde{\Delta }_{\text{SAS}}=\Delta _{\text{SAS}}/\sqrt{1-\sigma
_{0}^{2}}$. Thus, when the condition (\ref{CondiPpinExcitEH}) is satisfied,
an electron is excited to the $|\psi _{\text{p}}\rangle $ state at the
balanced point, and it makes a sudden transition to the $|\psi _{\text{sr}%
}\rangle $ state, and finally transforms smoothly into the $|\psi _{\text{s}%
}\rangle $ state, as depicted in FIG.\ref{FigExcitEH}. The sudden transition
would be smoothed out in a skyrmion-antiskyrmion excitation involving
several electrons simultaneously.

\subsection{Physical Densities}

It is interesting to investigate a single electron excitation and a single
hole excitation separately. For simplicity we analyze the SU(2) QH
ferromagnet. One-electron excited state $|\mathfrak{S}_{+}\rangle $ and
one-hole excited state $|\mathfrak{S}_{-}\rangle $ are given by%
\begin{equation}
|\mathfrak{S}_{+}\rangle =c_{\downarrow }^{\dagger }(0)\prod_{n}c_{\uparrow
}^{\dagger }(n)|0\rangle ,\mathbf{\quad }|\mathfrak{S}_{-}\rangle
=\prod_{n\neq 0}c_{\uparrow }^{\dagger }(n)|0\rangle ,  \label{ElectState}
\end{equation}%
where we have placed an electron or a hole at the momentum-zero state. Their
classical densities are 
\begin{align}
\rho _{\pm }^{\text{cl}}(m,n)=& \sum_{\sigma }\langle \mathfrak{S}_{\pm
}|c_{\sigma }^{\dagger }(m)c_{\sigma }(n)|\mathfrak{S}_{\pm }\rangle  \notag
\\
=& \delta _{mn}\pm \delta _{m0}\delta _{n0}.  \label{ClassDensiEHmn}
\end{align}%
In the coordinate space, based on formula (\ref{WeylOrderDensiX}) it reads 
\begin{equation}
\widehat{\rho }_{\pm }^{\text{cl}}(\mathbf{x})=\rho _{0}\left( 1\pm
2e^{-r^{2}/\ell _{B}^{2}}\right) ,  \label{DensiUnphys}
\end{equation}%
where $\rho _{0}$ is the electron density in the ground state. The hole
density becomes negative at the origin, 
\begin{equation}
\lim_{\mathbf{x}\rightarrow 0}\widehat{\rho }_{-}^{\text{cl}}(\mathbf{x}%
)=-\rho _{0}.  \label{FalseDensi}
\end{equation}%
There is nothing wrong with this mathematically since the electron cannot be
localized within the LLL. Nevertheless, we cannot accept this as a physical
quantity.

We recall that the wave function of one electron with the angular-momentum
zero in the LLL is%
\begin{equation}
\varphi _{0}(\mathbf{x})=\frac{1}{\sqrt{2\pi \ell _{B}^{2}}}e^{-r^{2}/4\ell
_{B}^{2}},
\end{equation}%
which leads to the density 
\begin{equation}
\delta \rho ^{\text{cl}}(\mathbf{x})=|\varphi _{0}(\mathbf{x})|^{2}=\rho
_{0}e^{-r^{2}/2\ell _{B}^{2}}  \label{PhysiDensiEH}
\end{equation}%
at $\nu =2\pi \ell _{B}^{2}\rho _{0}=1$. When we remove one electron from or
add one electron to the filled up-spin level, the density becomes%
\begin{equation}
\rho _{\pm }^{\text{cl}}(\mathbf{x})=\rho _{0}\pm \delta \rho ^{\text{cl}}(%
\mathbf{x}),  \label{DensiPhysi}
\end{equation}%
and the hole density satisfies 
\begin{equation}
\lim_{\mathbf{x}\rightarrow 0}\rho _{-}^{\text{cl}}(\mathbf{x})=0.
\end{equation}%
This behavior is what we expect for the physical density at the origin.

It is easy to see that the Fourier transformations of (\ref{DensiUnphys})
and (\ref{DensiPhysi}) are related as%
\begin{equation}
\rho _{\text{h}}^{\text{cl}}(\mathbf{q})=e^{-\ell _{B}^{2}\mathbf{q}^{2}/4}%
\widehat{\rho }_{\text{h}}^{\text{cl}}(\mathbf{q}).  \label{Naive2WeylClass}
\end{equation}%
This is precisely the relation (\ref{Naive2Weyl}) between the two types of
the densities. We conclude that the projected density represents a physical
quantity.

\subsection{Direct and Exchange Energies}

For simplicity we work still in the SU(2) QH ferromagnet. The electron-hole
pair excitation energy is\cite{Kallin91B}%
\begin{equation}
\delta E_{\text{pair}}=\langle \psi _{\text{s}}|H_{\text{C}}|\psi _{\text{s}%
}\rangle -E_{\text{g}}=2\epsilon _{\text{X}},
\end{equation}%
where%
\begin{equation}
\epsilon _{\text{X}}={V}_{0000}+\sum_{j\neq 0}V_{j00j}=\Delta _{\text{C}%
}^{0},
\end{equation}%
as is obtained by taking $d\rightarrow 0$ in (\ref{ParamXPM}) and (\ref%
{PairEnergEH}). It follows from (\ref{Vmnij}) that ${V}_{0000}$ is the
direct integral and $V_{j00j}$ ($j\neq 0$) is the exchange integral. Hence,
the Coulomb energy consists of the direct energy $\delta H_{\text{D}}^{\text{%
cl}}={V}_{0000}$ and the exchange energy $\delta H_{\text{X}}^{\text{cl}%
}=\sum_{j\neq 0}V_{j00j}$.

We can express them in familiar forms. First, the classical energy
associated with the density modulation (\ref{ClassDensiEHmn}) reads%
\begin{equation}
\delta H_{\text{D}}^{\text{cl}}=\sum_{mnij}{V}_{mnij}\delta \rho ^{\text{cl}%
}(m,n)\delta \rho ^{\text{cl}}(i,j)={V}_{0000}  \label{ClassEnergD}
\end{equation}%
with $\delta \rho ^{\text{cl}}(m,n)=\delta _{m0}\delta _{n0}$, which is
transformed into 
\begin{equation}
\delta H_{\text{D}}^{\text{cl}}={\frac{1}{2}}\int \!d^{2}xd^{2}y\;V(\mathbf{x%
}-\mathbf{y})\delta \rho ^{\text{cl}}(\mathbf{x})\delta \rho ^{\text{cl}}(%
\mathbf{y})
\end{equation}%
with $\delta \rho ^{\text{cl}}(\mathbf{x})$ being given by (\ref%
{PhysiDensiEH}). This is the direct energy of the excitation. We next remark
that the spin reads 
\begin{align}
S_{x}^{\text{cl}}(m,n)=& S_{y}^{\text{cl}}(m,n)=0,  \notag \\
S_{z}^{\text{cl}}(m,n)=& \frac{1}{2}\left( \delta _{mn}-\delta _{m0}\delta
_{n0}\right)
\end{align}%
both for the electron or hole excitation. Substituting them into the
exchange energy (\ref{SU4InvarMN}), or%
\begin{align}
H_{\text{X}}^{\text{cl}}=& -2\sum_{mnij}{V}_{mnij}\big[\sum_{a=xyz}S_{a}^{%
\text{cl}}(m,j)S_{a}^{\text{cl}}(i,n)  \notag \\
& \hspace{0.8in}+\frac{1}{4}\rho _{\pm }^{\text{cl}}(m,j)\rho _{\pm }^{\text{%
cl}}(i,n)\big],
\end{align}%
we reproduce $\delta H_{\text{X}}^{\text{cl}}=\sum_{j\neq 0}V_{j00j}$. In
the momentum space we have a more familiar expression, 
\begin{align}
H_{\text{X}}^{\text{cl}}=& \pi \int d^{2}p\,V_{\text{X}}(\mathbf{p})\big[%
\sum_{a=xyz}\widehat{S}_{a}^{\text{cl}}(-\mathbf{p})\widehat{S}_{a}^{\text{cl%
}}(\mathbf{p})  \notag \\
& \hspace{0.8in}+\frac{1}{4}\widehat{\rho }^{\text{cl}}(-\mathbf{p})\widehat{%
\rho }^{\text{cl}}(\mathbf{p})\big].
\end{align}%
Thus, the excitation energy of an electron or a hole presents us the
simplest example of the decomposition formula (\ref{CintoDX}) into the
direct and exchange energies.

\section{Skyrmions in Microscopic Theory}

We study skyrmions in a microscopic theory of the SU(2) QH ferromagnet, and
then extend the scheme to the SU(4) bilayer QH ferromagnet. We set $\ell
_{B}=1$ throughout in this section, where $\rho _{0}=1/2\pi $.

\subsection{SU(2) Skyrmions}

A skyrmion and an antiskyrmion are topological solitons in the nonlinear
sigma model, and characterized by the nonlinear sigma field (normalized spin
field), 
\begin{equation}
\mathcal{S}_{x}=\frac{\alpha x}{r^{2}+\alpha ^{2}},\quad \mathcal{S}_{y}=%
\frac{\mp \alpha y}{r^{2}+\alpha ^{2}},\quad \mathcal{S}_{z}=\frac{1}{2}-%
\frac{\alpha ^{2}}{r^{2}+\alpha ^{2}},  \label{SkyrmStepB}
\end{equation}%
when it carries topological charge $\pm 1$. Actual skyrmions are much more
complicated in the QH ferromagnet because the spin rotation modulates the
electron density due to the W$_{\infty }$(2) algebra. Nevertheless, a gross
feature remains as it is\cite{BookEzawa}.

Let $|\mathfrak{S}\rangle $ be the skyrmion state. Its spin field is given
by [see (\ref{WeylOrderDensiX})] 
\begin{equation}
\widehat{S}_{a}^{\text{cl}}(\mathbf{p})=\frac{1}{2\pi }\sum_{mn}^{\infty
}\langle m|e^{-i\mathbf{pX}}|n\rangle \langle \mathfrak{S}|S_{a}(m,n)|%
\mathfrak{S}\rangle .
\end{equation}%
Due to the formula%
\begin{align}
\langle m|& e^{i\mathbf{kX}}|m+n\rangle  \notag \\
& =\frac{\sqrt{m!}}{\sqrt{(m+n)!}}\left( \frac{k_{y}+ik_{x}}{\sqrt{2}}%
\right) ^{n}e^{-\mathbf{k}^{2}/4}L_{m}^{n}\left( \frac{\mathbf{k}^{2}}{2}%
\right) ,  \label{FormuXmn}
\end{align}%
we need to have $\langle \mathfrak{S}|S_{x,y}(m,n)|\mathfrak{S}\rangle
\varpropto \delta _{m,n\pm 1}$ and $\langle \mathfrak{S}|S_{z}(m,n)|%
\mathfrak{S}\rangle \varpropto \delta _{m,n}$ to be consistent with (\ref%
{SkyrmStepB}). Such a state is uniquely constructed as\cite{Fertig94B}%
\begin{equation}
|\mathfrak{S}_{\text{sky}}\rangle =\prod_{n=0}\xi ^{\dagger }(n)|0\rangle
\label{MicroSkyrmState}
\end{equation}%
with%
\begin{equation}
\xi ^{\dagger }(n)=u(n)c_{\downarrow }^{\dagger }(n)+v(n)c_{\uparrow
}^{\dagger }(n+1).
\end{equation}%
They satisfy the standard canonical commutation relations,%
\begin{equation}
\lbrack \xi (m),\xi ^{\dagger }(n)]=\delta _{mn},\quad \lbrack \xi (m),\xi
(n)]=0,
\end{equation}%
provided $u^{2}(n)+v^{2}(n)=1$. Note the state $|\mathfrak{S}_{\text{sky}%
}\rangle $ describes one hole state in the momentum-zero site when we set $%
u(n)=0$ and $v(n)=1$ for all $n$.

Similarly, the antiskymion state is given by\cite{Fertig94B}%
\begin{equation}
|\mathfrak{S}_{\text{asky}}\rangle =\prod_{n=0}\zeta ^{\dagger
}(n)c_{\downarrow }^{\dagger }(0)|0\rangle
\end{equation}%
with%
\begin{equation}
\zeta ^{\dagger }(n)=-u(n)c_{\downarrow }^{\dagger }(n+1)+v(n)c_{\uparrow
}^{\dagger }(n).
\end{equation}%
The state $|\mathfrak{S}_{\text{asky}}\rangle $ describes one electron
excited state in the momentum-zero site when we set $u(n)=0$ and $v(n)=1$
for all $n$. In what follow we only discuss the skyrmion case explicitly.

First we evaluate $\widehat{\rho }^{\text{cl}}(m,n)=\langle \mathfrak{S}%
|\rho (m,n)|\mathfrak{S}\rangle $ and others. The only nonvanishing
components are [see (\ref{AppCDXc}) in Appendix \ref{AppenDandE}]%
\begin{align}
\widehat{\rho }^{\text{cl}}(m,m)& =u^{2}(m)+v^{2}(m-1),  \notag \\
\widehat{S}_{z}^{\text{cl}}(m,m)& =-\frac{1}{2}\left[ u^{2}(m)-v^{2}(m-1)%
\right] ,  \notag \\
\widehat{S}_{x}^{\text{cl}}(m,m+1)& =\widehat{S}_{x}^{\text{cl}}(m+1,m)=-i%
\widehat{S}_{y}^{\text{cl}}(m,m+1)  \notag \\
& =i\widehat{S}_{y}^{\text{cl}}(m+1,m)=\frac{1}{2}u(m)v(m),
\label{MicroDensiMM}
\end{align}%
where we have set $v(-1)=0$. They are converted into the momentum space
based on the formula (\ref{WeylOrderDensiX}),%
\begin{align}
\widehat{\rho }^{\text{cl}}(\mathbf{k})=& \frac{1}{2\pi }e^{-k^{2}/4}%
\sum_{m=0}\left[ u^{2}(m)+v^{2}(m-1)\right] L_{m}\left( k^{2}/2\right) , 
\notag \\
\widehat{S}_{z}^{\text{cl}}(\mathbf{k})=& -\frac{1}{4\pi }%
e^{-k^{2}/4}\sum_{m=0}\left[ u^{2}(m)-v^{2}(m-1)\right] L_{m}\left(
k^{2}/2\right) ,  \notag \\
\widehat{S}_{x}^{\text{cl}}(\mathbf{k})=& \frac{-ik_{a}}{2\sqrt{2}\pi }%
e^{-k^{2}/4}\sum_{m=0}\frac{u(m)v(m)}{\sqrt{m+1}}L_{m}^{(1)}\left(
k^{2}/2\right) ,  \notag \\
\widehat{S}_{y}^{\text{cl}}(\mathbf{k})=& \frac{ik_{a}}{2\sqrt{2}\pi }%
e^{-k^{2}/4}\sum_{m=0}\frac{u(m)v(m)}{\sqrt{m+1}}L_{m}^{(1)}\left(
k^{2}/2\right)  \label{DensiUVk}
\end{align}%
after some calculation with the use of (\ref{FormuXmn}).

The electron number defference between the skyrmion state and the ground
state is 
\begin{align}
\delta Q=& \sum_{m=0}^{\infty }[\widehat{\rho }^{\text{cl}}(m,m)-1]  \notag
\\
=& u^{2}(0)-1+\sum_{m=1}^{\infty }\left[ u^{2}(m)-u^{2}(m-1)\right] =-1
\label{SkyrmChargMicro}
\end{align}%
by pairwise cancellations, where we have used $v^{2}(m)=1-u^{2}(m)$, $m\geq
0 $, and $v(-1)=0$. Hence the skyrmion excitation removes one electron from
the ground state. However, because the manipulation is too subtle, we give a
concrete calculation taking an explicit example in (\ref{MicroSkyrmCharg}).

As we have remarked before, it is necessary to construct physical densities
from the densities (\ref{DensiUVk}) by way of (\ref{Naive2WeylClass}). The
procedure is simply to replace $e^{-k^{2}/4}$ with $e^{-k^{2}/2}$ therein.
We make the Fourier transformation of the physical densities with the aid of
the formula [see formula (2.19.12.6) in Ref.\cite{PrudnikovIS2}]%
\begin{align}
\int_{0}^{\infty }{\!}kdk{\,}k^{m}& e^{-k^{2}/2}J_{m}(kr)L_{n}^{(m)}\left(
k^{2}/2\right)  \notag \\
=& \frac{1}{n!}\left( \frac{r^{2}}{2}\right) ^{n}r^{m}e^{-r^{2}/2},
\end{align}%
and 
\begin{align}
& \oint \!d\theta \,e^{i\mathbf{kx}}=2\pi J_{0}(kr),  \notag \\
& \frac{1}{2\pi i}\oint \!d\theta \,\frac{k_{j}}{k}e^{i\mathbf{kx}}=\frac{%
x_{j}}{r}J_{1}(kr),  \label{MathFormuA}
\end{align}%
where we have set $\mathbf{x}=(r\cos \theta ,r\sin \theta )$. The physical
densities reads 
\begin{align}
\rho ^{\text{cl}}(\mathbf{x})=& \frac{1}{2\pi }e^{-r^{2}/2}\sum_{m=0}\left[
u^{2}(m)+v^{2}(m-1)\right] \left( \frac{r^{2}}{2}\right) ^{n},  \notag \\
S_{z}^{\text{cl}}(\mathbf{x})=& -\frac{1}{2\pi }e^{-r^{2}/2}\sum_{m=0}\frac{%
u^{2}(m)-v^{2}(m-1)}{n!}\left( \frac{r^{2}}{2}\right) ^{n},  \notag \\
S_{x}^{\text{cl}}(\mathbf{x})=& \frac{x}{2\pi \sqrt{2}}e^{-r^{2}/2}\sum_{m=0}%
\frac{u(m)v(m)}{n!\sqrt{n+1}}\left( \frac{r^{2}}{2}\right) ^{n},  \notag \\
S_{y}^{\text{cl}}(\mathbf{x})=& -\frac{y}{2\pi \sqrt{2}}e^{-r^{2}/2}%
\sum_{m=0}\frac{u(m)v(m)}{n!\sqrt{n+1}}\left( \frac{r^{2}}{2}\right) ^{n}
\end{align}%
in the coordinate space.

In order to make a further analysis we make an anzats\cite{Fertig94B} on
functions $u(m)$ and $v(m)$,%
\begin{equation}
u^{2}(n)=\frac{\omega ^{2}}{n+1+\omega ^{2}},\quad v^{2}(n)=\frac{n+1}{%
n+1+\omega ^{2}}.  \label{SkyrmAnzatL}
\end{equation}%
The densities can be expressed in terms of the Kummer function $M(a;b;x)$, 
\begin{equation}
M(a;a+1;x)=a\sum_{n=0}^{\infty }\frac{x^{n}}{(n+a)n!},
\end{equation}%
as 
\begin{align}
\rho (\mathbf{x})=& \frac{1}{2\pi }-\frac{1}{2\pi }\frac{1}{\omega ^{2}+1}%
\hspace*{0.5mm}e^{-\frac{1}{2}r^{2}}M(\omega ^{2};\omega ^{2}+2;r^{2}/2), 
\notag \\
S_{z}(\mathbf{x})=& \frac{1}{4\pi }-\frac{1}{4\pi }e^{-\frac{1}{2}%
r^{2}}M(\omega ^{2};\omega ^{2}+1;r^{2}/2)  \notag \\
& -\frac{1}{4\pi }\frac{\omega ^{2}}{\omega ^{2}+1}e^{-\frac{1}{2}%
r^{2}}M(\omega ^{2}+1;\omega ^{2}+2;r^{2}/2),  \notag \\
S_{x}(\mathbf{x})=& \frac{1}{4\pi }\frac{\sqrt{2}\omega x}{\omega ^{2}+1}e^{-%
\frac{1}{2}r^{2}}M(\omega ^{2}+1;\omega ^{2}+2;r^{2}/2),  \notag \\
S_{y}(\mathbf{x})=& \frac{1}{4\pi }\frac{-\sqrt{2}\omega y}{\omega ^{2}+1}%
e^{-\frac{1}{2}r^{2}}M(\omega ^{2}+1;\omega ^{2}+2;r^{2}/2),
\end{align}%
where we have used the formula (13.4.3) of Ref.\cite{Abramowitz} to derive $%
\rho (\mathbf{x})$. The electron number of the skyrmion excitation is 
\begin{align}
\delta Q=& \int \!d^{2}x\;\left[ \rho (\mathbf{r})-\frac{1}{2\pi }\right] 
\notag \\
=& \frac{-1}{\omega ^{2}+1}\int_{0}^{\infty }e^{-z}M(\omega ^{2};\omega
^{2}+2;z)\hspace*{0.25mm}dz=-1,  \label{MicroSkyrmCharg}
\end{align}%
where the last equality is checked via order by order integration with
respect to $z$. There is no ambiguity in this derivation contrary to that in
(\ref{SkyrmChargMicro}).

The skyrmion with the anzats (\ref{SkyrmAnzatL}) has a peculiar feature. It
is reduced to a hole for $\omega =0$, where the density $\rho (\mathbf{x})$
approaches the ground-state value exponentially fast. However, for all $%
\omega \neq 0$, with the use of the formula (13.1.4) of Ref.\cite{Abramowitz}%
, we find 
\begin{equation}
\rho (\mathbf{x})\rightarrow \frac{1}{2\pi }\left( 1-\frac{2\alpha ^{2}}{%
r^{4}}\right) ,\qquad \text{as}\quad r\rightarrow \infty ,
\end{equation}%
where we have set $\alpha ^{2}=2\omega ^{2}$. Furthermore, we find 
\begin{align}
& S_{x}(\mathbf{x})\rightarrow \frac{1}{2\pi }\frac{\alpha x}{r^{2}},\quad
S_{y}(\mathbf{x})\rightarrow \frac{-1}{2\pi }\frac{\alpha y}{r^{2}},  \notag
\\
& S_{z}(\mathbf{x})\rightarrow \frac{1}{2\pi }\left( \frac{1}{2}-\frac{%
\alpha ^{2}}{r^{2}}\right) ,\qquad \text{as}\quad r\rightarrow \infty .
\end{align}%
They agree with the asymptotic behaviors of the density and the spin field
of a sufficiently large skyrmion we discuss later: See (\ref{DensiModul}).
It is easy to see that the number of the flipped spin diverges for all $%
\omega \neq 0$. Consequently, small skyrmions cannot be discussed based on
this anzats.

\subsection{SU(4) Skyrmions}

The generalization to the SU(4) bilayer system is straightforward. The
ground state is given by the up-spin bonding state. The skyrmion state is
given by 
\begin{equation}
|\mathfrak{S}_{\text{sky}}\rangle =\prod_{n=0}\xi ^{\dagger }(n)|0\rangle ,
\label{MicroSkyrmSU4}
\end{equation}%
where%
\begin{align}
\xi ^{\dagger }(n)=& u_{\text{r}}(n)A_{\downarrow }^{\dagger }(n)+u_{\text{p}%
}(n)A_{\upharpoonleft }^{\dagger }(n)  \notag \\
& +u_{\text{s}}(n)B_{\downarrow }^{\dagger }(n)+v(n)B_{\uparrow }^{\dagger
}(n+1)
\end{align}%
with constraint, $u_{\text{r}}^{2}(n)+u_{\text{p}}^{2}(n)+u_{\text{s}%
}^{2}(n)+v^{2}(n)=1$. The hole state $B_{\uparrow }(0)|$B$\uparrow \rangle $
is given by setting $v(n)=1$ and $u_{\text{s}}(n)=u_{\text{p}}(n)=u_{\text{r}%
}(n)=0$ for all $n$.

There are three types of antiskyrmions,%
\begin{align}
|\mathfrak{S}_{\text{asky}}^{\text{s}}\rangle =& \prod_{n=0}\zeta
_{s}^{\dagger }(n)B_{\downarrow }^{\dag }(0)|0\rangle ,  \notag \\
|\mathfrak{S}_{\text{asky}}^{\text{p}}\rangle =& \prod_{n=0}\zeta
_{p}^{\dagger }(n)A_{\uparrow }^{\dagger }(0)|0\rangle ,  \notag \\
|\mathfrak{S}_{\text{asky}}^{\text{r}}\rangle =& \prod_{n=0}\zeta
_{r}^{\dagger }(n)A_{\downarrow }^{\dagger }(0)|0\rangle ,
\label{MicroASkyrmSU4}
\end{align}%
with 
\begin{align}
\zeta _{s}^{\dagger }(n)=& u_{\text{r}}(n)A_{\downarrow }^{\dagger }(n)+u_{%
\text{p}}(n)A_{\uparrow }^{\dagger }(n)  \notag \\
& -u_{\text{s}}(n)B_{\downarrow }^{\dagger }(n+1)+v(n)B_{\uparrow }^{\dagger
}(n+1),  \notag \\
\zeta _{p}^{\dagger }(n)=& u_{\text{r}}(n)A_{\downarrow }^{\dagger }(n)-u_{%
\text{p}}(n+1)A_{\uparrow }^{\dagger }(n)  \notag \\
& +u_{\text{s}}(n)B_{\downarrow }^{\dagger }(n)+v(n)B_{\uparrow }^{\dagger
}(n+1),  \notag \\
\zeta _{r}^{\dagger }(n)=& -u_{\text{r}}(n)A_{\downarrow }^{\dagger
}(n+1)+u_{\text{p}}(n)A_{\uparrow }^{\dagger }(n)  \notag \\
& +u_{\text{s}}(n)B_{\downarrow }^{\dagger }(n)+v(n)B_{\uparrow }^{\dagger
}(n+1).  \label{ASkyrmGenerSU4}
\end{align}%
They are reduced to three different electron excited states $B_{\downarrow
}^{\dag }(0)|$B$\uparrow \rangle $, $A_{\uparrow }^{\dagger }(0)|$B$\uparrow
\rangle $ and $A_{\downarrow }^{\dagger }(0)|$B$\uparrow \rangle $ when we
set $v(n)=1$ and $u_{\text{s}}(n)=u_{\text{p}}(n)=u_{\text{r}}(n)=0$ for all 
$n$: See FIG.\ref{FigLevelEH}.

There are important SU(2) limits of SU(4) skyrmions. When we set $u_{\text{p}%
}(n)=u_{\text{r}}(n)=0$, $|\mathfrak{S}_{\text{sky}}^{\text{s}}\rangle $ and 
$|\mathfrak{S}_{\text{asky}}^{\text{s}}\rangle $ describe a skyrmion and an
antiskyrmion where only spins are excited. We call them the spin-skyrmion
and the spin-antiskyrmion. Similarly, when we set $u_{\text{s}}(n)=u_{\text{r%
}}(n)=0$, $|\mathfrak{S}_{\text{sky}}^{\text{p}}\rangle $ and $|\mathfrak{S}%
_{\text{asky}}^{\text{p}}\rangle $ describe a skyrmion and an antiskyrmion
where only pseudospins are excited. We call them the ppin-skyrmion and the
ppin-antiskyrmion.

It is a dynamical problem which skyrmion-antiskymion pairs are excited
thermally. As we have shown, as far as electron-hole pairs are concerned,
only pseudospins are excited at the balanced point unless the tunneling gap
is too large, while only spins are excited at the monolayer point. This is
the case also for skyrmion-antiskyrmion pair excitations. However, in
general, all components are excited to lower the total energy, which leads
to genuine SU(4) skyrmions. Contrary to the case of the electron-hole limit,
the transition from the ppin-skyrmion at the balanced point ($\sigma _{0}=0$%
) to the spin-skyrmion at the monolayer point ($\sigma _{0}=1$) will occur
continuously via a genuine SU(4) skyrmion since the matrix elements of the
total Hamiltonian $\widehat{H}$ between various skyrmion states are
nonvanishing.

\section{Effective Hamiltonians}

It is very hard to calculate the skyrmion excitation energy with use of the
microscopic states (\ref{MicroSkyrmSU4}) and (\ref{MicroASkyrmSU4}) since
they involve infinitely many functions $u_{\text{s}}(n)$, $u_{\text{p}}(n)$, 
$u_{\text{r}}(n)$ and $v(n)$. In this paper we construct the effective
theory by making the derivative expansion of the Hamiltonian. Thus, strictly
speaking, our approximation is good only for large skyrmions. Nevertheless,
its application even to small skyrmions would present us invaluable results
otherwise unavailable. We wish to develop a microscopic theory in a future
work.

Our analysis is based on the decomposition formula (\ref{CintoDX}) into the
direct and exchange energy terms. In what follows we represent the classical
density $\rho ^{\text{cl}}(\mathbf{x})$ simply by $\rho (\mathbf{x})$ since
we use only classical fields.

We first identify the SU(4)-invariant direct Coulomb term as the self-energy,%
\begin{equation}
H_{\text{self}}={\frac{1}{2}}\int \!d^{2}xd^{2}y\,\rho (\mathbf{x})V^{+}(%
\mathbf{x}-\mathbf{y})\rho (\mathbf{y}).  \label{SelfEnerg}
\end{equation}%
We then make the derivative expansion of the SU(4)-invariant exchange term $%
H_{\text{X}}^{+}$. We rewrite it as 
\begin{align}
H_{\text{X}}^{+}=& -\frac{1}{2}\int \!d^{2}xd^{2}z\,V_{\text{X}}^{+}(\mathbf{%
z})\big[\widehat{\mathbf{I}}(\mathbf{x}+\mathbf{z})\widehat{\mathbf{I}}(%
\mathbf{x})  \notag \\
& \hspace{0.8in}+\frac{1}{8}\widehat{\rho }(\mathbf{x}+\mathbf{z})\widehat{%
\rho }(\mathbf{x})\big].  \label{HamilDeriv}
\end{align}%
Since $V_{\text{X}}(\mathbf{z})$ is short ranged,%
\begin{align}
V_{\text{X}}^{\pm }(\mathbf{x})=& \frac{1}{2\pi }\int \!d^{2}p\,e^{i\mathbf{%
px}}V_{\text{X}}^{\pm }(\mathbf{p})  \notag \\
=& V(\mathbf{x})\left( 1\pm e^{-|\mathbf{x}|d/\ell _{B}^{2}}\right) e^{-%
\mathbf{x}^{2}/2\ell _{B}^{2}},
\end{align}%
it is a good approximation to make the Taylor expansion of $\widehat{\mathbf{%
I}}(\mathbf{x}+\mathbf{z})$ and $\widehat{\rho }(\mathbf{x}+\mathbf{z})$\ to
the nontrivial lowest order of $\mathbf{z}$,%
\begin{equation}
\widehat{\mathbf{I}}(\mathbf{x}+\mathbf{z})\widehat{\mathbf{I}}(\mathbf{x})=%
\widehat{\mathbf{I}}(\mathbf{x})\widehat{\mathbf{I}}(\mathbf{x})-\frac{1}{2}%
\sum_{ij}z_{i}z_{j}\partial _{i}\widehat{\mathbf{I}}(\mathbf{x})\partial _{j}%
\widehat{\mathbf{I}}(\mathbf{x}),
\end{equation}%
where a partial integration is understood in the integrand of (\ref%
{HamilDeriv}). Equivalently, we make the momentum expansion of the potential
(\ref{KerneX}),%
\begin{equation}
V_{\text{X}}^{\pm }(\mathbf{p})=V_{\text{X}}^{\pm }(0)-8\pi \ell
_{B}^{4}J_{s}^{\pm }\mathbf{p}^{2}+O(\mathbf{p}^{4}),  \label{PotenExpanPM}
\end{equation}%
where 
\begin{equation}
V_{\text{X}}^{\pm }(0)={2}\ell _{B}^{2}\left[ 1\pm e^{d^{2}/2\ell _{B}^{2}}%
\text{erfc}\left( d/\sqrt{2}\ell _{B}\right) \right] \Delta _{\text{C}}^{0},
\label{StiffPM}
\end{equation}%
and%
\begin{equation}
J_{s}^{\pm }=\frac{1}{2}\left( J_{s}\pm J_{s}^{d}\right) ,
\label{IntroStiffPM}
\end{equation}%
with%
\begin{align}
J_{s}& ={\frac{1}{8\pi }}\Delta _{\text{C}}^{0},  \label{IntroStiffS} \\
{\frac{J_{s}^{d}}{J_{s}}}& =-\sqrt{\frac{2}{\pi }}{\frac{d}{\ell _{B}}}%
+\left( 1+{\frac{d^{2}}{\ell _{B}^{2}}}\right) e^{d^{2}/2\ell _{B}^{2}}\text{%
erfc}\left( d/\sqrt{2}\ell _{B}\right) .  \label{IntroStiffD}
\end{align}%
The zeroth order term in $\mathbf{p}^{2}$ is proportional to the integral%
\begin{equation}
\int \!d^{2}x\,\left[ \widehat{\mathbf{I}}(\mathbf{x})\widehat{\mathbf{I}}(%
\mathbf{x})+\frac{1}{8}\widehat{\rho }(\mathbf{x})\widehat{\rho }(\mathbf{x})%
\right] =\frac{1}{4\pi \ell _{B}^{2}}\int \!d^{2}x\,\widehat{\rho }(\mathbf{x%
}),  \label{CasimIdentA}
\end{equation}%
which is a Casimir invariant obtained by integrating the relation (\ref%
{CasimInX}). Note that the star product becomes an ordinary product within
the integrand. We may neglect the term, because it represents the total
number of electrons and is fixed to the ground-state value in excitations of
skyrmion-antiskyrmion pairs. Consequently, we obtain 
\begin{equation}
H_{\text{X}}^{+\text{eff}}=\frac{2J_{s}^{+}}{\rho _{\Phi }^{2}}\int
\!d^{2}x\,\left[ \partial _{k}\widehat{\mathbf{I}}(\mathbf{x})\partial _{k}%
\widehat{\mathbf{I}}(\mathbf{x})+\frac{1}{2N}\partial _{k}\widehat{\rho }(%
\mathbf{x})\partial _{k}\widehat{\rho }(\mathbf{x})\right]
\label{ExchaHamilL}
\end{equation}%
as the effective Hamiltonian. Here and hereafter the summation over the
repeated index in $\partial _{k}$ is understood.

We next derive the effective Hamiltonian from the SU(4)-noninvariant terms $%
H_{\text{C}}^{-}$. The zeroth order term in $\mathbf{p}^{2}$ yields the
capacitance energy as the leading term,%
\begin{equation}
H_{\text{cap}}={2}\pi \ell _{B}^{2}\epsilon _{\text{cap}}\int \!d^{2}x\,%
\widehat{P}_{z}(\mathbf{x})\widehat{P}_{z}(\mathbf{x}),  \label{CapacFormu}
\end{equation}%
where we have used the identity (\ref{CasimIdentA}). The capacitance energy (%
$\varpropto \epsilon _{\text{cap}}$) consists of two terms; the one ($%
\varpropto \epsilon _{\text{D}}^{-}$) arising from the direct interaction $%
H_{\text{D}}^{-}$, which is the standard capacitance energy of a condenser
made of two planes with separation $d$, and the other ($\varpropto \epsilon
_{\text{X}}^{-}$) from the exchange interaction $H_{\text{X}}^{-}$. The
exchange effect makes the capacitance energy quite small for a small layer
separation\cite{NoteCapac}. We note that our capacitance formula (\ref%
{CapacParamDX}) is different from the one assumed in some literature\cite%
{Moon95B,Demler99L}.

Collecting all the first order terms in $\mathbf{p}^{2}$ from the exchange
Hamiltonian (\ref{TruncX}) we obtain 
\begin{align}
\mathcal{H}_{\text{X}}^{\text{SU(4)}}& =\frac{1}{2}\pi ^{2}\ell _{B}^{4}J_{s}%
\left[ \partial _{k}\widehat{\rho }\right] ^{2}  \notag \\
& +{4}\pi ^{2}\ell _{B}^{4}J_{s}^{d}\left( \left[ \partial _{k}\widehat{S}%
_{a}\right] ^{2}+\left[ \partial _{k}\widehat{P}_{a}\right] ^{2}+\left[
\partial _{k}\widehat{R}_{ab}\right] ^{2}\right)  \notag \\
& +{8}\pi ^{2}\ell _{B}^{4}J_{s}^{-}\left( \left[ \partial _{k}\widehat{S}%
_{a}\right] ^{2}+\left[ \partial _{k}\widehat{P}_{z}\right] ^{2}+\left[
\partial _{k}\widehat{R}_{az}\right] ^{2}\right) ,  \label{ExchaSU4}
\end{align}%
where the summation over repeated indices $a$ and $b$ is understood.

It is worthwhile to take two important limits of $\mathcal{H}_{\text{X}}^{%
\text{SU(4)}}$. When all electrons are moved to the front layer, by setting $%
\Psi =(\psi ^{\text{f}\uparrow },\psi ^{\text{f}\downarrow },0,0)$, the
nonvanishing elements are $\widehat{S}_{a}=\widehat{S}_{a}^{\text{f}}$ and $%
\widehat{R}_{az}=\widehat{S}_{a}^{\text{f}}$ in (\ref{ExchaSU4}), and we
find 
\begin{equation}
\mathcal{H}_{\text{X}}^{\text{spin}}=\frac{2J_{s}}{\rho _{\Phi }^{2}}\sum %
\left[ \partial _{k}\widehat{S}_{a}^{\text{f}}\right] ^{2}
\label{HamilSpinFerro}
\end{equation}%
for the spin-ferromagnet. Similarly, when the spin degree of freedom is
frozen, the nonvanishing elements are $\widehat{P}_{a}=\widehat{P}%
_{a}^{\uparrow }$ and $\widehat{R}_{az}=\widehat{P}_{a}^{\uparrow }$ in (\ref%
{ExchaSU4}), and we find 
\begin{equation}
\mathcal{H}_{\text{X}}^{\text{ppin}}=\frac{2J_{s}^{d}}{\rho _{\Phi }^{2}}%
\sum_{a=x,y}[\partial _{k}\widehat{P}_{a}^{\uparrow }(\mathbf{x})]^{2}+\frac{%
2J_{s}}{\rho _{\Phi }^{2}}[\partial _{k}\widehat{P}_{z}^{\uparrow }(\mathbf{x%
})]^{2}  \label{HamilPpinFerro}
\end{equation}%
for the pseudospin-ferromagnet.

To discuss the Goldstone modes we may set $\widehat{\rho }(\mathbf{x})=\rho
_{0}$ since the ground state is robust against the density fluctuation. By
setting $\widehat{S}_{a}^{\text{f}}=\rho _{\Phi }\mathcal{S}_{a}$, (\ref%
{HamilSpinFerro}) becomes an O(3) nonlinear sigma model describing the
spin-ferromagnet Hamiltonian\cite{Sondhi93B} with the spin stiffness $J_{s}$%
. By setting $\widehat{P}_{a}^{\uparrow }=\rho _{\Phi }\mathcal{P}_{a}$, (%
\ref{HamilPpinFerro}) becomes an anisotropic O(3) nonlinear sigma model
describing the pseudospin-ferromagnet Hamiltonian\cite{Moon95B} with the
interlayer stiffness $J_{s}^{d}$.

\section{Semiclassical Analysis}

We use bosonic variables to describe coherent excitations such as spin and
pseudospin textures. In so doing we introduce the composite-boson (CB) field%
\cite{Girvin87L,Read89L}. The CB theory of QH ferromagnets is formulated as
follows\cite{Ezawa99L}. The CB field $\phi ^{\sigma }(\mathbf{x})$ is
defined by making a singular phase transformation to the electron field $%
\psi _{\sigma }(\mathbf{x})$, 
\begin{equation}
\phi ^{\sigma }(\mathbf{x})=e^{-ie\Theta (\mathbf{x})}\psi _{\sigma }(%
\mathbf{x}),  \label{BareCB}
\end{equation}%
where the phase field $\Theta (\mathbf{x})$ attaches one flux quantum to
each electron via the relation,%
\begin{equation}
\varepsilon _{ij}\partial _{i}\partial _{j}\Theta (\mathbf{x})=\Phi _{\text{D%
}}\rho (\mathbf{x}).  \label{PhaseField}
\end{equation}%
We then introduce the normalized CB field $n^{\sigma }(\mathbf{x})$ by 
\begin{equation}
\phi ^{\sigma }(\mathbf{x})=\phi (\mathbf{x})n^{\sigma }(\mathbf{x}),
\label{NormaCB}
\end{equation}%
where $\phi ^{\dag }(\mathbf{x})\phi (\mathbf{x})=\sum_{\sigma }\psi
_{\sigma }^{\dag }(\mathbf{x})\psi _{\sigma }(\mathbf{x})=\rho (\mathbf{x)}$%
, and the $4$-component field $n^{\sigma }(\mathbf{x})$ obeys the constraint 
$\sum_{\sigma }n^{\sigma \dagger }(\mathbf{x})n^{\sigma }(\mathbf{x})=1$:
Such a field is the CP$^{3}$ field\cite{DAdda78NPB}. The isospin operators
are expressed as%
\begin{align}
S_{a}(\mathbf{x})=& \rho (\mathbf{x)}\mathcal{S}_{a}(\mathbf{x}),\quad 
\mathcal{S}_{a}(\mathbf{x})=\frac{1}{2}\mathbf{n}^{\dag }(\mathbf{x})\tau
_{a}^{\text{spin}}\mathbf{n}(\mathbf{x}),  \notag \\
P_{a}(\mathbf{x})=& \rho (\mathbf{x)}\mathcal{P}_{a}(\mathbf{x}),\quad 
\mathcal{P}_{a}(\mathbf{x})=\frac{1}{2}\mathbf{n}^{\dag }(\mathbf{x})\tau
_{a}^{\text{ppin}}\mathbf{n}(\mathbf{x}),  \label{SpinPpin}
\end{align}%
and so on, with $\mathbf{n}=(n^{\text{f}\uparrow },n^{\text{f}\downarrow
},n^{\text{b}\uparrow },n^{\text{b}\downarrow })^{T}$. They are%
\begin{align}
\mathcal{S}_{x}^{\text{g}}& =0,\quad \mathcal{S}_{y}^{\text{g}}=0,\quad 
\mathcal{S}_{z}^{\text{g}}=\frac{1}{2},  \notag \\
\mathcal{P}_{x}^{\text{g}}& =\frac{1}{2}\sqrt{1-\sigma _{0}^{2}},\quad 
\mathcal{P}_{y}^{\text{g}}=0,\quad \mathcal{P}_{z}^{\text{g}}=\frac{1}{2}%
\sigma _{0},  \label{GrounSP}
\end{align}%
and $\mathcal{R}_{ab}^{\text{g}}=0$ except for $\mathcal{R}_{zx}^{\text{g}}=%
\mathcal{P}_{x}^{\text{g}}$ and $\mathcal{R}_{zz}^{\text{g}}=\mathcal{P}%
_{z}^{\text{g}}$ in the ground state.

We investigate charged excitations at $\nu =1$. Charged excitations are
topological solitons in incompressible QH liquids. To describe them we
introduce the dressed CB\ field by\cite{Ezawa99L,BookEzawa} 
\begin{equation}
\varphi ^{\alpha }(\mathbf{x})=e^{-\mathcal{A}(\mathbf{x})}\phi ^{\alpha }(%
\mathbf{x}),  \label{DressCB}
\end{equation}%
where $\phi ^{\alpha }(\mathbf{x})$ is given by (\ref{BareCB}) and $\mathcal{%
A}(\mathbf{x})$ is given by 
\begin{equation}
\mathcal{A}(\mathbf{x})=\int \!d^{2}y\;\ln \left( {\frac{|\mathbf{x}-\mathbf{%
y}|}{2\ell _{B}}}\right) \rho (\mathbf{y})-|z|^{2}.  \label{ConstAfield}
\end{equation}%
The kinetic Hamiltonian (\ref{HamilKinem}) is rewritten as 
\begin{equation}
H_{\text{K}}={\frac{1}{2M}}\sum_{\alpha }\int d^{2}x\Phi ^{\ddag }(\mathbf{x}%
)(\mathcal{D}_{x}-i\mathcal{D}_{y})(\mathcal{D}_{x}+i\mathcal{D}_{y})\Phi (%
\mathbf{x}),  \label{HamilCoher}
\end{equation}%
where $\Phi (\mathbf{x})$ is the four-component dressed CB field with $\Phi
^{\ddag }=\Phi ^{\dagger }e^{2\mathcal{A}}$, and $\mathcal{D}_{j}=-i\hbar
\partial _{j}+\hbar (\varepsilon _{jk}+\delta _{jk})\partial _{k}\mathcal{A}(%
\mathbf{x})$.

The LLL condition follows from the kinetic Hamiltonian (\ref{HamilCoher}),%
\begin{equation}
\left( \mathcal{D}_{x}+i\mathcal{D}_{y}\right) \Phi (\mathbf{x})|\mathfrak{S}%
\rangle =-{\frac{i\hbar }{\ell _{B}}}{\frac{\partial }{\partial z^{\ast }}}%
\Phi (\mathbf{x})|\mathfrak{S}\rangle =0,  \label{PpinCondiLLL}
\end{equation}%
where $z=(x+iy)/2\ell _{B}$. It implies that the $N$-body wave function is
analytic and symmetric in $N$ variables, 
\begin{equation}
\mathfrak{S}_{\text{CB}}[z]=\langle 0|\Phi (\mathbf{x}_{1})\cdots \Phi (%
\mathbf{x}_{N})|\mathfrak{S}\rangle .
\end{equation}%
It is easy to verify\cite{Ezawa99L,BookEzawa} that the electron wave
function is $\mathfrak{S}[\mathbf{x}]=\mathfrak{S}_{\text{CB}}[z]\mathfrak{S}%
_{\text{Laughlin}}[\mathbf{x}]$, where $\mathfrak{S}_{\text{Laughlin}}[%
\mathbf{x}]$ is the Laughlin wave function\cite{Laughlin83L}.

The analysis is quite simple when the function is factorized, $\mathfrak{S}_{%
\text{CB}}[z]=\prod_{r}\mathfrak{S}(z_{r})$. Then it follows that $\mathfrak{%
S}(z)=\langle \Phi (\mathbf{x})\rangle $ and that $\langle \mathbf{n}(%
\mathbf{x})\rangle =\mathfrak{S}(z)/|\mathfrak{S}(z)|$. The lightest
topological soliton is described by the nontrivial simplest wave function $(%
\mathfrak{S}^{\text{B}\uparrow },\mathfrak{S}^{\text{B}\downarrow },%
\mathfrak{S}^{\text{A}\uparrow },\mathfrak{S}^{\text{A}\downarrow
})=(z,\kappa _{s},\kappa _{p},\kappa _{r})$, which we call the SU(4)
skyrmion. Thus, a skyrmion is characterized by its shape parameters $\kappa
_{s}$, $\kappa _{p}$ and $\kappa _{r}$ representing how it is excited to
energy levels B$\downarrow $, A$\uparrow $ and A$\downarrow $, respectively
[FIG.\ref{FigLevelEH}]. In terms of the layer CP$^{3}$ field it reads%
\begin{equation}
\left( 
\begin{array}{@{\,}r}
n^{\text{f}\uparrow }(\mathbf{x}) \\ 
n^{\text{f}\downarrow }(\mathbf{x}) \\ 
n^{\text{b}\uparrow }(\mathbf{x}) \\ 
n^{\text{b}\downarrow }(\mathbf{x})%
\end{array}%
\right) =C(z)\left( 
\begin{array}{@{\,}c}
z\sqrt{1+\sigma _{0}}+\kappa _{p}\sqrt{1-\sigma _{0}} \\ 
\kappa _{s}\sqrt{1+\sigma _{0}}+\kappa _{r}\sqrt{1-\sigma _{0}} \\ 
z\sqrt{1-\sigma _{0}}-\kappa _{p}\sqrt{1+\sigma _{0}} \\ 
\kappa _{s}\sqrt{1-\sigma _{0}}-\kappa _{r}\sqrt{1+\sigma _{0}}%
\end{array}%
\right) ,  \label{SU4Skyrm}
\end{equation}%
with the normalization factor $C(z)=1/\sqrt{2(z^{2}+\kappa ^{2})}$ with $%
\kappa ^{2}=\kappa _{s}^{2}+\kappa _{p}^{2}+\kappa _{r}^{2}$. When $\kappa
=\kappa _{s}$, $\kappa _{p}=\kappa _{r}=0$, it is reduced to 
\begin{equation}
\left( 
\begin{array}{@{\,}r}
n^{\text{f}\uparrow }(\mathbf{x}) \\ 
n^{\text{f}\downarrow }(\mathbf{x}) \\ 
n^{\text{b}\uparrow }(\mathbf{x}) \\ 
n^{\text{b}\downarrow }(\mathbf{x})%
\end{array}%
\right) _{\text{spin}}=C(z)\left( 
\begin{array}{@{\,}r}
z\sqrt{1+\sigma _{0}} \\ 
\kappa \sqrt{1+\sigma _{0}} \\ 
z\sqrt{1-\sigma _{0}} \\ 
\kappa \sqrt{1-\sigma _{0}}%
\end{array}%
\right) ,  \label{SkyrmSwaveL}
\end{equation}%
which describes a spin texture reversing only spins: This is identified with
the microscopic spin-skyrmion in (\ref{MicroSkyrmSU4}). When $\kappa =\kappa
_{p}$, $\kappa _{s}=\kappa _{r}=0$, it is reduced to 
\begin{equation}
\left( 
\begin{array}{@{\,}r}
n^{\text{f}\uparrow }(\mathbf{x}) \\ 
n^{\text{f}\downarrow }(\mathbf{x}) \\ 
n^{\text{b}\uparrow }(\mathbf{x}) \\ 
n^{\text{b}\downarrow }(\mathbf{x})%
\end{array}%
\right) _{\text{ppin}}=C(z)\left( 
\begin{array}{@{\,}r}
z\sqrt{1+\sigma _{0}}+\kappa \sqrt{1-\sigma _{0}} \\ 
0 \\ 
z\sqrt{1-\sigma _{0}}-\kappa \sqrt{1+\sigma _{0}} \\ 
0%
\end{array}%
\right) ,  \label{SkyrmPwaveL}
\end{equation}%
which describes a pseudospin texture reversing only pseudospins: This is
identified with the microscopic ppin-skyrmion in (\ref{MicroSkyrmSU4}).

A skyrmion excitation modulates the density around it, $\rho _{0}\rightarrow
\rho _{\text{sky}}(\mathbf{x})$, according to the soliton equation\cite%
{Ezawa99L},%
\begin{equation}
{\frac{1}{4\pi }}\mathbf{\nabla }^{2}\ln \rho _{\text{sky}}(\mathbf{x})-\rho
_{\text{sky}}(\mathbf{x})+\rho _{0}=J_{\text{sky}}(\mathbf{x}),
\label{SolitEqSkyrm}
\end{equation}%
which follows from the LLL condition (\ref{PpinCondiLLL}): $J_{\text{sky}}(%
\mathbf{x})$ is the topological (Pontryagin number) density, which is
calculated as%
\begin{equation}
J_{\text{sky}}(\mathbf{x})={\frac{1}{\pi }}{\frac{4(\kappa \ell _{B})^{2}}{%
[r^{2}+4(\kappa \ell _{B})^{2}]^{2}}}  \label{SkyrmPontr}
\end{equation}%
for the SU(4) skyrmion (\ref{SU4Skyrm}). The soliton equation is solved
iteratively, and the first order term is%
\begin{equation}
\delta \rho _{\text{sky}}(\mathbf{x})\simeq -J_{\text{sky}}(\mathbf{x})=-{%
\frac{1}{\pi }}{\frac{4(\kappa \ell _{B})^{2}}{[r^{2}+4(\kappa \ell
_{B})^{2}]^{2}}.}  \label{DensiModul}
\end{equation}%
This is good for a very smooth skyrmion ($\kappa \gg 1$).

The antiskyrmion configuration is related to the skyrmion configuration by%
\begin{equation}
\begin{tabular}{lll}
$\mathcal{S}_{x}^{\text{asky}}=\mathcal{S}_{x}^{\text{sky}},\quad $ & $%
\mathcal{S}_{y}^{\text{asky}}=-\mathcal{S}_{y}^{\text{sky}},\quad $ & $%
\mathcal{S}_{z}^{\text{asky}}=\mathcal{S}_{z}^{\text{sky}},$ \\ 
$\mathcal{P}_{x}^{\text{asky}}=\mathcal{P}_{x}^{\text{sky}},$ & $\mathcal{P}%
_{y}^{\text{asky}}=-\mathcal{P}_{y}^{\text{sky}},$ & $\mathcal{P}_{z}^{\text{%
asky}}=\mathcal{P}_{z}^{\text{sky}}.$%
\end{tabular}%
\end{equation}%
A skyrmion (antiskyrmion) induces the modulation of the spin and the
pseudospin,%
\begin{align}
\delta S_{a}^{\text{(a)sky}}(\mathbf{x})& =\rho _{\text{(a)sky}}(\mathbf{x})%
\mathcal{S}_{a}^{\text{(a)sky}}(\mathbf{x})-\rho _{0}\mathcal{S}_{a}^{\text{g%
}},  \notag \\
\delta P_{a}^{\text{(a)sky}}(\mathbf{x})& =\rho _{\text{(a)sky}}(\mathbf{x})%
\mathcal{P}_{a}^{\text{(a)sky}}(\mathbf{x})-\rho _{0}\mathcal{P}_{a}^{\text{g%
}},  \label{SkyAntiSky}
\end{align}%
where $\mathcal{S}_{a}^{\text{g}}$ and $\mathcal{P}_{a}^{\text{g}}$ are the
ground-state values (\ref{GrounSP}), and%
\begin{align}
\rho _{\text{sky}}(\mathbf{x})=& \rho _{0}+\delta \rho _{\text{sky}}(\mathbf{%
x}),  \notag \\
\rho _{\text{asky}}(\mathbf{x})=& \rho _{0}-\delta \rho _{\text{sky}}(%
\mathbf{x}),
\end{align}%
where $\delta \rho _{\text{sky}}(\mathbf{x})$ is given by (\ref{DensiModul}).

It is actually the energy of a skyrmion-antiskyrmion pair, 
\begin{equation}
\Delta _{\text{pair}}=\frac{1}{2}(E_{\text{sky}}+E_{\text{asky}}),
\label{PairEnerg}
\end{equation}%
that is observed experimentally. We estimate the excitation energy $E_{\text{%
sky}}=\langle \mathfrak{S}|\widehat{H}|\mathfrak{S}\rangle $ of one skyrmion%
\cite{NoteSingleSkyrm}. For simplicity we set $\kappa _{r}=0$ in the SU(4)
skyrmion (\ref{SU4Skyrm}). This approximation reduces the validity of some
of our results. Indeed, we have found that the mixing between the excitation
modes to the s-level and the r-level lowers the energy of the electron-hole
pair state [FIG.\ref{FigExcitEH}]. Nevertheless, we use this approximation
to reveal an essential physics of SU(4) skyrmions, because otherwise various
formulas become too complicated to handle with. By an essential physics we
mean a continuous transformation of the SU(4) skyrmion from the
ppin-skyrmion limit to the spin-skyrmion limit in contrast to the case of
the electron-hole excitation.

Thus we study a skyrmion parametrized by two shape parameters $\kappa _{s}$
and $\kappa _{p}$ with $\kappa ^{2}=\kappa _{s}^{2}+\kappa _{p}^{2}$. We
calculate the SU(4) generators (\ref{SpinPpin}) for this field configuration,%
\begin{align}
\mathcal{S}_{x}^{\text{sky}}& =\frac{\alpha _{s}x}{r^{2}+\alpha ^{2}},\quad 
\mathcal{S}_{y}^{\text{sky}}=\frac{-\alpha _{s}y}{r^{2}+\alpha ^{2}},\quad 
\mathcal{S}_{z}^{\text{sky}}=\frac{1}{2}-\frac{\alpha _{s}^{2}}{r^{2}+\alpha
^{2}},  \notag \\
\mathcal{P}_{x}^{\text{sky}}& =\frac{\cos \beta }{2}-\frac{\alpha _{p}x\sin
\beta +\alpha _{p}^{2}\cos \beta }{r^{2}+\alpha ^{2}},\quad \mathcal{P}_{y}^{%
\text{sky}}=\frac{\alpha _{p}y}{r^{2}+\alpha ^{2}},  \notag \\
\mathcal{P}_{z}^{\text{sky}}& =\frac{\sin \beta }{2}+\frac{\alpha _{p}x\cos
\beta -\alpha _{p}^{2}\sin \beta }{r^{2}+\alpha ^{2}},
\label{IpinFieldSkyrm}
\end{align}%
and similar expressions for $\mathcal{R}_{ab}^{\text{sky}}$, where $\alpha
_{s}=2\kappa _{s}\ell _{B}$, $\alpha _{p}=2\kappa _{p}\ell _{B}$, $\alpha
^{2}=4(\kappa \ell _{B})^{2}$, $\cos \beta =\sqrt{1-\sigma _{0}^{2}}$ and $%
\sin \beta =\sigma _{0}$.

The skyrmion energy consists of the Coulomb energy, the Zeeman energy and
the pseudo-Zeeman energy. The Coulomb energy consists of the self-energy,
the capacitance energy and the exchange energy.

The dominant one is the self-energy (\ref{SelfEnerg}), 
\begin{equation}
E_{\text{self}}={\frac{1}{2}}\int \!d^{2}xd^{2}y\,\delta \rho _{\text{sky}}(%
\mathbf{x})V^{+}(\mathbf{x}-\mathbf{y})\delta \rho _{\text{sky}}(\mathbf{y}).
\end{equation}%
After a straightforward calculation we find%
\begin{equation}
E_{\text{self}}=\frac{1}{8\kappa }E_{\text{C}}^{0}\int
z^{2}[K_{1}(z)]^{2}\left( 1+e^{-\frac{d}{2\ell }\frac{z}{\kappa }}\right) dz.
\label{DirecEnergSkyrm}
\end{equation}%
See (\ref{AppSelfEnerg}) in Appendix \ref{AppenCouloEnerg}. It depends only
on the total skyrmion scale $\kappa =\sqrt{\kappa _{s}^{2}+\kappa _{p}^{2}}$%
, since the Coulomb term $H_{\text{D}}^{+}$ depends only on the total
density $\delta \rho _{\text{sky}}(\mathbf{x})$.

The capacitance energy is given by (\ref{CapacFormu}), or 
\begin{equation}
H_{\text{cap}}={2}\pi \ell _{B}^{2}\epsilon _{\text{cap}}\int
\!d^{2}x\;\delta P_{z}^{\text{sky}}(\mathbf{x})\delta P_{z}^{\text{sky}}(%
\mathbf{x}).  \label{CapacFormC1}
\end{equation}%
It is calculated in (\ref{AppCouloM}) in Appendix \ref{AppenCouloEnerg}. The
leading term is 
\begin{equation}
E_{\text{cap}}\simeq \frac{1}{2}(1-\sigma _{0}^{2})\epsilon _{\text{cap}}N_{%
\text{ppin}}(\kappa _{p}),  \label{EnergSkyrmDm}
\end{equation}%
where $N_{\text{ppin}}(\kappa _{p})$ is the number of flipped pseudospins to
be defined by (\ref{NumbePpinX}).

In evaluating the exchange energy (\ref{ExchaSU4}), we set $\rho (\mathbf{x}%
)=\rho _{0}$ since the skyrmion charge is spread over a large domain. (See
Appendix B for the result without making the approximation.) Using (\ref%
{IpinFieldSkyrm}) we obtain%
\begin{equation}
E_{\text{X}}=4\pi \left[ J_{s}^{+}-\frac{1}{3}J_{s}^{-}\left\{ \frac{\kappa
_{p}^{2}}{\kappa ^{2}}-\frac{\kappa _{s}^{2}}{\kappa ^{2}}\left( 1+2\frac{%
\kappa _{s}^{2}}{\kappa ^{2}}\right) \sigma _{0}^{2}\right\} \right] .
\label{IpinExchaEnerg}
\end{equation}%
It contains shape parameters explicitly via the SU(4)-noninvariant term. The
SU(4)-invariant part of (\ref{ExchaSU4}) is the SU(4) nonlinear sigma model
yielding a topological invariant value $4\pi J_{s}^{+}$. In the
spin-skyrmion limit it is reduced to%
\begin{equation}
E_{\text{X}}^{\text{spin}}=4\pi \left( J_{s}^{+}+J_{s}^{-}\sigma
_{0}^{2}\right) ,  \label{SpinXEnerg}
\end{equation}%
which yields the well-known formula\cite{Sondhi93B}, $E_{\text{X}}^{\text{%
spin}}=4\pi J_{s}$, at the monolayer point ($\sigma _{0}\rightarrow 1$). In
the ppin-skyrmion limit it is reduced to%
\begin{equation}
E_{\text{X}}^{\text{ppin}}=4\pi (J_{s}^{+}-\frac{1}{3}J_{s}^{-})=\frac{4\pi 
}{3}\left( J_{s}+2J_{s}^{d}\right) ,  \label{PpinXEnerg}
\end{equation}%
which is independent of the imbalance parameter $\sigma _{0}$.

The Zeeman energy of one skyrmion is 
\begin{equation}
E_{\text{Z}}=-\Delta _{\text{Z}}\int \!d^{2}x\,\delta S_{z}^{\text{sky}}(%
\mathbf{x})\equiv N_{\text{spin}}(\kappa _{s})\Delta _{\text{Z}},
\label{SkyrmZeemaBL}
\end{equation}%
where $N_{\text{spin}}$ is the number of flipped spins. Neglecting the term
which is cancelled out in a skyrmion-antiskyrmion pair excitation due to the
relations (\ref{SkyAntiSky}), we obtain 
\begin{equation}
N_{\text{spin}}(\kappa _{s})=\rho _{0}\int \!d^{2}x\,\frac{4(\kappa _{s}\ell
_{B})^{2}}{r^{2}+4(\kappa \ell _{B})^{2}}\simeq \kappa _{s}^{2}N_{\xi }
\label{NumbeSpin}
\end{equation}%
with%
\begin{equation}
N_{\xi }=2\ln \left( 1+{\frac{\xi ^{2}}{4\ell _{B}^{2}}}\right) ,
\label{FuncN}
\end{equation}%
where the divergence has been cut off at $r\simeq \kappa \xi $ with a
typical coherence length $\xi $.

The pseudo-Zeeman energy (\ref{PseudZeemaX}) is 
\begin{equation}
E_{\text{PZ}}=-\Delta _{\text{SAS}}\int \!d^{2}x\left[ \delta P_{x}^{\text{%
sky}}(\mathbf{x})+{\frac{\sigma _{0}\delta P_{z}^{\text{sky}}(\mathbf{x})}{%
\sqrt{1-\sigma _{0}^{2}}}}\right] .  \label{PpinEnerg}
\end{equation}%
We extract the terms which are not cancelled out in a skyrmion-antiskyrmion
pair excitation. It is equal to%
\begin{equation}
E_{\text{PZ}}=\frac{N_{\text{ppin}}(\kappa _{p})}{\sqrt{1-\sigma _{0}^{2}}}%
\Delta _{\text{SAS}},  \label{SkyrmPZeema}
\end{equation}%
where $N_{\text{ppin}}$ is the number of flipped pseudospins,%
\begin{equation}
N_{\text{ppin}}(\kappa _{p})\simeq \kappa _{p}^{2}N_{\xi }.
\label{NumbePpinX}
\end{equation}%
We have cut off the divergence as in (\ref{NumbeSpin}).

\begin{figure}[t]
\begin{center}
\includegraphics[width=0.34\textwidth]{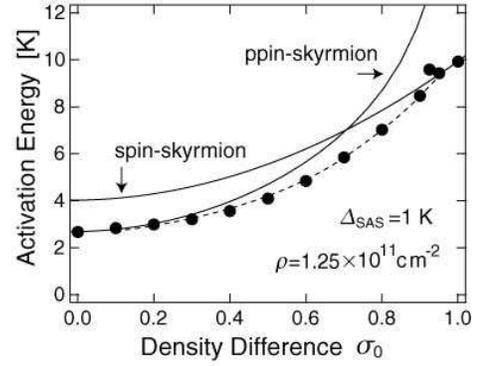}
\end{center}
\caption{ The skyrmion energy is estimated as a function of the imbalanced
parameter $\protect\sigma _{0}$ with sample parameters $\Delta _{\text{SAS}%
}=1$K, and $\protect\rho _{0}=1.2\times 10^{11}$/cm$^{2}$. The experimental
data are taken from Terasawa et al.\protect\cite{Terasawa04}. Solid lines
show the energies of a ppin-skyrmion and a spin-skyrmion, normalized to the
data at the balanced point and the monolayer point, respectively. The
experimental data would be explained by the excitation of a genuine SU(4)
skyrmion. }
\label{FigBLSkyTh}
\end{figure}

The total excitation energy is 
\begin{equation}
E_{\text{sky}}=E_{\text{X}}+E_{\text{self}}+E_{\text{cap}}+E_{\text{Z}}+E_{%
\text{PZ}}  \label{SkyrmEnergFunct}
\end{equation}%
with (\ref{IpinExchaEnerg}), (\ref{DirecEnergSkyrm}), (\ref{EnergSkyrmDm}), (%
\ref{SkyrmZeemaBL}) and (\ref{SkyrmPZeema}). Skyrmion parameters $\kappa
_{s} $ and $\kappa _{p}$ with $\kappa ^{2}=\kappa _{s}^{2}+\kappa _{p}^{2}$
are to be determined by minimizing the excitation energy (\ref%
{SkyrmEnergFunct}). According to an examination of (\ref{SkyrmEnergFunct})
presented in Appendix \ref{AppenGenerSkyrm}, provided%
\begin{equation}
\frac{4\pi J_{s}^{-}}{3}>\left( \frac{1}{2}\epsilon _{\text{cap}}-\Delta _{%
\text{Z}}+\Delta _{\text{SAS}}\right) \kappa ^{2}N_{\xi },
\label{CondiPpinExcitSky}
\end{equation}%
ppin-skyrmions are excited at the balanced point ($\sigma _{0}=0$): See (\ref%
{CondiPpinExcit}). Then, it evolves continuously into a spin-skyrmion at the
monolayer point ($\sigma _{0}=1$) via a generic skyrmion ($\kappa _{s}\kappa
_{p}\neq 0$). As we have stated, to simplify calculations we have set $%
\kappa _{r}=0$ in the SU(4) skyrmion (\ref{SU4Skyrm}). When we allow a
skyrmion to be excited into the r-level ($\kappa _{r}\neq 0$), a genuine
SU(4) skyrmions ($\kappa _{s}\kappa _{p}\kappa _{r}\neq 0$) would be excited
except at the monolayer point, as illustrated in FIG.\ref{FigBLSkyTh}.
Compare this with FIG.\ref{FigExcitEH}.

We examine the condition (\ref{CondiPpinExcitSky}) numerically. When we
adopt typical values of sample parameters ($\rho _{0}=1.2\times 10^{11}$/cm$%
^{2}$ and $d=231$\AA ), we find the capacitance (\ref{CapacParamDX}) to be $%
\epsilon _{\text{cap}}\simeq 132$K while the exchange-energy difference to
be $4\pi J_{s}^{-}/3\simeq 5.4$K. The condition is hardly satisfied. Here,
we question the validity of the standard identification of the layer
separation, $d=d_{B}+d_{W}$, where $d_{W}\simeq 200$\AA\ is the width of a
quantum well and $d_{B}\simeq 31$\AA\ is the separation of the two quantum
wells. In this identification the electron cloud is assumed be localized in
the center of each quantum well. However, it is a dynamical problem. Let us
minimize the ppin-skyrmion energy as a function of $d$. The minimum is found
to be achieved at $d<d_{B}$. Namely, the energy increases monotonously for $%
d>d_{B}$. Then, it would be reasonable to use $d=d_{B}$ as the layer
separation to estimate the energy of the ppin-skyrmion. When we choose $%
d\simeq 31$\AA\ we find $\epsilon _{\text{cap}}\simeq 4.5$K. Then, the
condition (\ref{CondiPpinExcitSky}) is satisfied in some parameter regions,
where ppin-skyrmions are excited.

However, the condition (\ref{CondiPpinExcitSky}) cannot be taken literally,
because it rules out excitations of large ppin-skyrmions ($\kappa ^{2}N_{\xi
}\gg 1$) against experimental indications\cite%
{Murphy94L,SawadaX03PE,Terasawa04}. It seems that the exchange energy has
been underestimated by making the derivative expansion. It is necessary to
go beyond the present approximation to fully understand the problem.

We proceed to estimate the energies of a spin-skyrmion and a ppin-skyrmion
in all range of $\sigma _{0}$, because their estimation is much more
reliable than that of a generic SU(4) skyrmion. We use sample parameters $%
\rho _{0}=1.2\times 10^{11}$/cm$^{2}$, $d=31$\AA\ and $\Delta _{\text{SAS}%
}=1 $K, and normalize the energy to the experimental data\cite{Terasawa04}
at the monolayer point for a spin-skyrmion and at the balanced point for a
ppin-skyrmion. Recall that the absolute value of the activation energy
cannot be determined theoretically because they depends essentially on
samples.

First, we calculate the energy of a spin-skyrmion as a function of $\sigma
_{0}$ by setting $\kappa =\kappa _{s}$ and $\kappa _{p}=0$ in (\ref%
{SkyrmEnergFunct}). The skyrmion scale $\kappa $, to be determined by
minimizing this, depends on $\sigma _{0}$ very weakly. We obtain $N_{\text{%
spin}}(\kappa )\simeq 3.5$ as in the monolayer QH ferromagnet\cite%
{Ezawa99L,BookEzawa}. We have depicted the excitation energy $E_{\text{sky}%
}^{\text{spin}}$ in FIG.\ref{FigBLSkyTh}, where it is normalized to the data
at the monolayer point.

Next, we calculate the energy of a ppin-skyrmion as a function of $\sigma
_{0}$ by setting $\kappa =\kappa _{p}$ and $\kappa _{s}=0$ in (\ref%
{SkyrmEnergFunct}). There is an important remark. The effective tunneling
gap diverges as $\sigma _{0}\rightarrow 1$ unless $\Delta _{\text{SAS}}=0$.
Furthermore, the minimum pseudospin flip is $N_{\text{ppin}}=1$ for a
skyrmion-antiskyrmion pair. Hence, the pair excitation energy (\ref%
{PairEnerg}) diverges as%
\begin{equation}
\Delta _{\text{pair}}\rightarrow \frac{\Delta _{\text{SAS}}}{\sqrt{1-\sigma
_{0}^{2}}},
\end{equation}%
and ppin-skyrmions are not excited at the monolayer point ($\sigma
_{0}\rightarrow 1$). We have depicted the excitation energy $E_{\text{sky}}^{%
\text{ppin}}$ in FIG.\ref{FigBLSkyTh}, where it is normalized to the data at
the monolayer point. In so doing we have used the full expression (\ref%
{AppCouloM}) for the capacitance energy $E_{\text{cap}}$ since other terms
are as important as (\ref{EnergSkyrmDm}) for a ppin-skyrmion of an ordinary
size.

\section{Activation Energy Anomaly}

\label{SecMurph}

We have studied how one skyrmion evolves continuously from the balanced
point ($\sigma _{0}=0$)\ to the monolayer point ($\sigma _{0}=1$)\ by
changing its shape. It is important how to distinguish various shapes of
skyrmions experimentally. As is well known, as the sample is tilted, the
activation energy of a spin-skyrmion increases due to the Zeeman energy. On
the contrary, the activation energy of a ppin-skyrmion decreases due to the
loss of the exchange energy. Thus, the tilted-field method provides us with
a remarkable experimental method\cite{Murphy94L,SawadaX03PE} to reveal the
existence of various shapes of a skyrmion in bilayer QH systems.

As the sample is tilted, the parallel magnetic field $B_{\parallel }$ is
penetrated between the two layers. We take the symmetric gauge generalized as%
\begin{equation}
\mathbf{A}=\left( \frac{1}{2}B_{\perp }y+B_{\Vert }z,-\frac{1}{2}B_{\perp
}x,0\right) ,  \label{GaugeAzZero}
\end{equation}%
where the two layers are placed at $z=\pm d/2$. In the kinetic Hamiltonian (%
\ref{HamilKinem}) the covariant momentum is now different between the two
layers, 
\begin{align}
D_{x}=& -i\hbar {\frac{\partial }{\partial x}}+{\frac{\hbar }{2\ell _{B}^{2}}%
}y+\frac{eB_{\Vert }d}{2}\tau _{z}^{\text{ppin}},  \notag \\
D_{y}=& -i\hbar {\frac{\partial }{\partial y}}-{\frac{\hbar }{2\ell _{B}^{2}}%
}x.
\end{align}%
Consequently, the LLL condition (\ref{PpinCondiLLL}) is modified as%
\begin{align}
(\mathcal{D}_{x}+& i\mathcal{D}_{y})\Phi (\mathbf{x})|\mathfrak{S};B_{\Vert
}\rangle  \notag \\
=& -{\frac{i\hbar }{\ell _{B}}}\left( {\frac{\partial }{\partial z^{\ast }}+}%
\frac{i}{2}\delta _{\text{m}}\ell _{B}\tau _{z}^{\text{ppin}}\right) \Phi (%
\mathbf{x})|\mathfrak{S};B_{\Vert }\rangle =0  \label{ParalCondiLLL}
\end{align}%
with%
\begin{equation}
\delta _{\text{m}}={\frac{edB_{\Vert }}{\hbar }}.  \label{MisFitPre}
\end{equation}%
We may solve the LLL condition (\ref{ParalCondiLLL}) for the one-body wave
function $\mathfrak{S}(\mathbf{x};B_{\Vert })=\langle \mathbf{x}|\mathfrak{S}%
;B_{\Vert }\rangle $\ as%
\begin{equation}
\mathfrak{S}(\mathbf{x};B_{\Vert })=\exp \left( {-}\frac{i}{2}\delta _{\text{%
m}}\tau _{z}^{\text{ppin}}x\right) \mathfrak{S}(z).  \label{AlphaWave}
\end{equation}%
Accordingly, the skyrmion configuration acquires different phase factors
between the two layers, 
\begin{equation}
\mathbf{n}(\mathbf{x};B_{\Vert })=\exp \left( {-}\frac{i}{2}\delta _{\text{m}%
}\tau _{z}^{\text{ppin}}x\right) \mathbf{n}(\mathbf{x};0),
\label{PpinSkyrmParal}
\end{equation}%
where $\mathbf{n}(\mathbf{x};0)$\ is the configuration (\ref{SU4Skyrm}) in
the absence of the parallel magnetic field. Various isospin fields are given
by (\ref{SpinPpin}) with this CP$^{3}$ field.

We first consider the balanced configuration ($\sigma _{0}=0$), where we
assume ppin-skyrmions are excited [FIG.\ref{FigBLSkyTh}]. The excitation
energy at $B_{\Vert }=0$ is given by, 
\begin{equation}
E_{\text{sky}}^{\text{ppin}}(0)=E_{\text{X}}^{\text{ppin}}(0)+E_{\text{self}%
}+E_{\text{cap}}+N_{\text{ppin}}\Delta _{\text{SAS}}{,}  \label{EnergBalan}
\end{equation}%
where $E_{\text{X}}^{\text{ppin}}(0)$ is the exchange energy (\ref%
{PpinXEnerg}) in the absence of the parallel magnetic field. We analyze how
the exchange Hamiltonian (\ref{HamilPpinFerro}) is affected by the parallel
magnetic field. Subtracting the ground-state energy we easily deduce the $%
B_{\Vert }$ dependence of the excitation energy,%
\begin{equation}
E_{\text{X}}^{\text{ppin}}(B_{\Vert })=E_{\text{X}}^{\text{ppin}}(0)+\Delta
E_{\text{X}}^{\text{ppin}}(B_{\Vert })  \label{ExchaEnergLoss}
\end{equation}%
with%
\begin{equation}
\Delta E_{\text{X}}^{\text{ppin}}(B_{\Vert })=-2J_{s}^{d}\delta _{m}^{2}\int
\!d^{2}x\,\left( \mathcal{P}_{z}^{\text{sky}}(\mathbf{x})\right) ^{2},
\label{ExchaHamilParal}
\end{equation}%
where $\mathcal{P}_{z}^{\text{sky}}$ is the skyrmion pseudospin component in
the absence of the parallel magnetic field and given by (\ref{IpinFieldSkyrm}%
).

\begin{figure}[t]
\begin{center}
\includegraphics[width=0.49\textwidth]{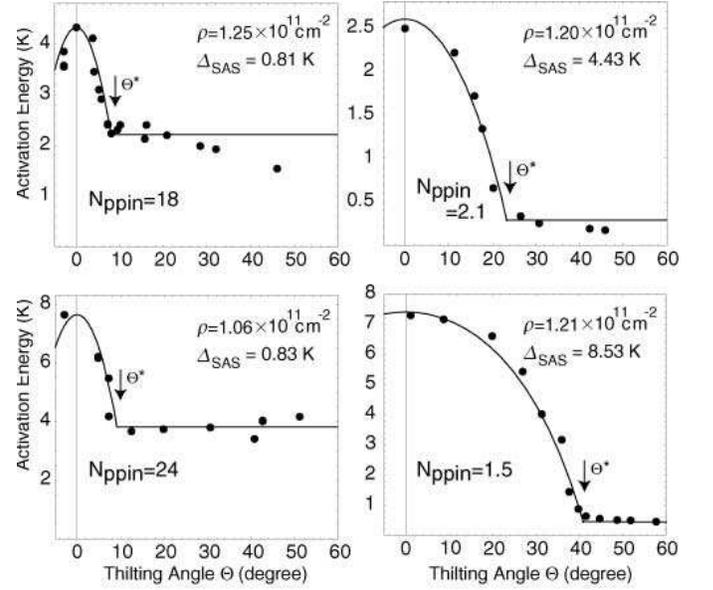}
\end{center}
\caption{ The activation energy at $\protect\nu =1$ is plotted as a function
of the tilting angle $\Theta $ in several samples with different tunneling
gaps $\Delta _{\text{SAS}}$. It shows a rapid decrease towards the critical
angle $\Theta ^{\ast }$, and then becomes almost flat. The data are taken
from Murphy et al.\protect\cite{Murphy94L}. They are well fitted by the
theoretical formula {(\protect\ref{TheorMurph})}, where the activation
energy decreases in the commensurate phase and becomes flat in the
incommensurate phase. To fit the data, we have adjusted the activation
energy at $\Theta =0$ with the experimental value, and assumed that the
number of flipped pseudospins $N_{\text{ppin}}$ is constant for all values
of the tilting angle.}
\label{FigSheena}
\end{figure}

We note that $\Delta E_{\text{X}}^{\text{ppin}}(B_{\Vert })$ is proportional
to the capacitance energy (\ref{CapacFormu}). Thus the leading order term is
found to be%
\begin{equation}
\Delta E_{\text{X}}^{\text{ppin}}(B_{\Vert })\simeq -\frac{2\pi
d^{2}J_{s}^{d}}{\ell _{B}^{2}}N_{\text{ppin}}\tan ^{2}\Theta ,
\label{PpinDecre}
\end{equation}%
where $\tan \Theta =B_{\Vert }/B_{\perp }$ and $N_{\text{ppin}}$ is the
number of pseudospins flipped around the skyrmion. All other terms in (\ref%
{EnergBalan}) are unaffected by the parallel magnetic field\cite{BookEzawa}.
Thus, the excitation energy decreases as the tilting angle $\Theta $
increases. The rate of the decrease depends on the number $N_{\text{ppin }}$%
of flipped pseudospins and the amount of the penetrated magnetic field $%
B_{\Vert }$.

It has been shown\cite{Moon95B,BookEzawa} that, when the parallel magnetic
field $B_{\Vert }$ increases more than a certain critical point, the phase
transition occurs in the bilayer QH system: It is the
commensurate-incommensurate transition point $B_{\Vert }^{\ast }$. In the
incommensurate phase the penetrated magnetic field is not increased more
than $B_{\Vert }^{\ast }$, because the excess magnetic field $\Delta
B_{\Vert }^{\ast }=B_{\Vert }^{\ast }-B_{\Vert }^{\ast }$ is eaten up to
create penetrated sine-Gordon vortices between the layers. Hence, the
activation energy becomes flat for $\Theta >\Theta ^{\ast }$, where $\tan
\Theta ^{\ast }=B_{\Vert }^{\ast }/B_{\perp }$.

Hence, from (\ref{ExchaEnergLoss}) and (\ref{PpinDecre}) the excitation
energy is%
\begin{equation}
E_{\text{sky}}^{\text{ppin}}(B_{\parallel })=E_{\text{X}}^{\text{ppin}%
}(0)+E_{\text{self}}+E_{\text{cap}}+N_{\text{ppin}}\Delta _{\text{SAS}%
}^{\Theta }{,}  \label{TheorMurph}
\end{equation}%
where we have absorbed $\Delta E_{\text{X}}^{\text{ppin}}(B_{\parallel })$
into the effective tunneling gap,%
\begin{equation}
\Delta _{\text{SAS}}^{\Theta }=\left\{ 
\begin{array}{ll}
\Delta _{\text{SAS}}-(2\pi d^{2}J_{s}^{d}/\ell _{B}^{2})\tan ^{2}\Theta & 
\text{for}\quad \Theta <\Theta ^{\ast } \\[3mm] 
\Delta _{\text{SAS}}-(2\pi d^{2}J_{s}^{d}/\ell _{B}^{2})\tan ^{2}\Theta
^{\ast } & \text{for}\quad \Theta >\Theta ^{\ast }%
\end{array}%
\right. .  \label{EffecTunne}
\end{equation}%
At each angle $\Theta $, in principle, it is necessary to minimize the total
excitation energy (\ref{TheorMurph}) and determine the scale parameter $%
\kappa $. However, since the energy change (\ref{EffecTunne}) is quite small
compared with the total energy, we may treat it as a small perturbation.
Namely, we may assume that the flipped pseudospin number $N_{\text{ppin}}$
is a constant independent of $\Theta $. This reminds us of experimental data%
\cite{Melinte99L,Kumada00L}, where the flipped spin number $N_{\text{spin}}$
is nearly constant by tilting the sample in the excitation of spin-skyrmions.

We have fitted the data due to Murphy et al.\cite{Murphy94L} in FIG.\ref%
{FigSheena} by assuming an appropriate flipped pseudospin number $N_{\text{%
ppin}}$ per one skyrmion-antiskyrmion pair. We summarize the result in a
table. 
\begin{equation}
\begin{tabular}{||l|l|l|l|l||}
\hline\hline
sample & A & B & C & D \\ \hline
$\Delta _{\text{SAS}}$ & $0.81$ & $0.86$ & $4.43$ & $8.53$ \\ \hline
$N_{\text{ppin}}$ & $18$ & $24$ & $2.1$ & $1.5$ \\ \hline\hline
\end{tabular}
\label{TableSheena}
\end{equation}%
It takes a large value in samples with $\Delta _{\text{SAS}}<1$ K but take a
small value in samples with $\Delta _{\text{SAS}}>4$ K. Recall spin-skyrmion
excitations in the monolayer QH system, where the flipped spin number
remains small when the Zeeman energy is moderate but becomes quite large
when the Zeeman energy is almost zero.

We proceed to analyze excitations of SU(4) skyrmions (\ref{PpinSkyrmParal})
with (\ref{SU4Skyrm}) in imbalanced configuration, where the exchange
Hamiltonian is given by (\ref{ExchaSU4}). After some calculation we find
that the exchange-energy loss $\Delta E_{\text{X}}^{\text{SU(4)}%
}(B_{\parallel })$ is again proportional to the capacitance energy, and the
leading order term is 
\begin{equation}
\Delta E_{\text{X}}(B_{\parallel })\simeq -\frac{2\pi d^{2}J_{s}^{d}}{\ell
_{B}^{2}}(1-\sigma _{0}^{2})N_{\text{ppin}}\tan ^{2}\Theta ,
\end{equation}%
which is reduced to (\ref{PpinDecre}) in the balanced point.

Hence, in the commensurate phase ($\Theta <\Theta _{\sigma }^{\ast }$) the
excitation energy turns out to be%
\begin{align}
E_{\text{sky}}(B_{\parallel })=& E_{\text{X}}(0)+E_{\text{self}}+E_{\text{cap%
}}{+}N_{\text{spin}}\sqrt{1+\tan ^{2}\Theta }\Delta _{\text{Z}}  \notag \\
& +N_{\text{ppin}}\Delta _{\text{BAB}}^{\Theta }  \label{TheorTeras}
\end{align}%
with%
\begin{equation}
\Delta _{\text{BAB}}^{\Theta }={\frac{1}{\sqrt{1-\sigma _{0}^{2}}}}\Delta _{%
\text{SAS}}-\frac{2\pi d^{2}J_{s}^{d}}{\ell _{B}^{2}}(1-\sigma _{0}^{2})\tan
^{2}\Theta .  \label{TheorTerasDelta}
\end{equation}%
In the incommensurate phase ($\Theta >\Theta _{\sigma }^{\ast }$) it is
given by this formula by replacing $\tan ^{2}\Theta $ with $\tan ^{2}\Theta
_{\sigma }^{\ast }$. Here, the commensurate-incommensurate transition point
increases slowly as\cite{Hanna01B}%
\begin{equation}
\Theta _{\sigma }^{\ast }=(1-\sigma _{0}^{2})^{-1/4}\Theta ^{\ast },
\label{SigmaCICx}
\end{equation}%
as the imbalance parameter $\sigma _{0}$ increases.

\begin{figure}[t]
\begin{center}
\includegraphics[width=0.49\textwidth]{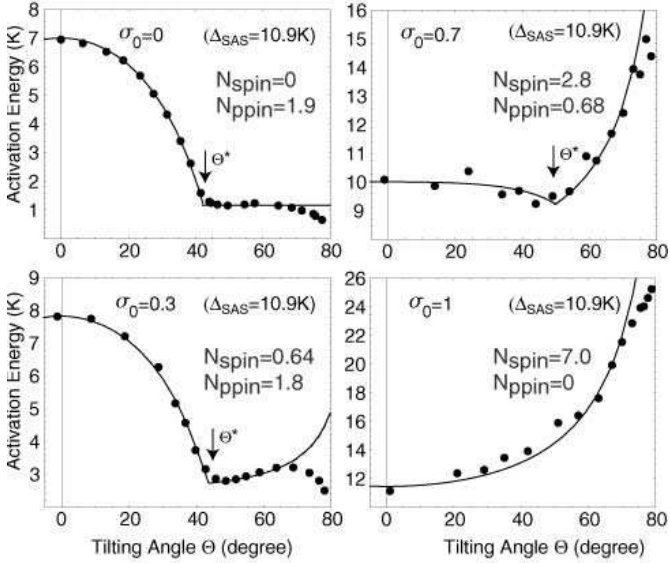}
\end{center}
\caption{ The activation energy at $\protect\nu =1$ is plotted as a function
of $\Theta $ in one sample ($\Delta _{\text{SAS}}=11$ K, $\protect\rho %
_{0}=0.6\times 10^{11}$/cm$^{2}$) with different imbalance parameter $%
\protect\sigma _{0}$. It shows a decrease towards the critical angle $\Theta
^{\ast }$, and then begin to increase for $\protect\sigma _{0}=0.3$ and $%
\protect\sigma _{0}=0.7$. The data are taken from Sawada et al.\protect\cite%
{SawadaX03PE}. To fit the data by the theoretical formula (\protect\ref%
{TheorTeras}), we have adjusted the activation energy at $\Theta =0$ with
the experimental value, and assumed that the flipped spin number $N_{\text{%
ppin}}$ and the flipped pseudospin number $N_{\text{ppin}}$ are constant for
all values of the tilting angle.}
\label{FigTerasawa}
\end{figure}

Experiments have been carried out by Sawada et al.\cite{SawadaX03PE} in
bilayer samples (FIG.\ref{FigTerasawa}), where the activation energy was
measured at $\nu =1$ by controlling both the tilting angle $\Theta $ and the
imbalance parameter $\sigma _{0}$. Their data are interpreted based on the
theoretical result (\ref{TheorTeras}) as follows. We focus on the behavior
of the activation energy $E_{\text{sky}}$ by changing the tilting angle at
each fixed imbalance parameter. The relevant term in (\ref{TheorTeras}) is%
\begin{align}
\Delta E_{\text{sky}}(\Theta )=& N_{\text{spin}}\sqrt{1+\tan ^{2}\Theta }%
\Delta _{\text{Z}}^{0}  \notag \\
& -N_{\text{ppin}}\frac{2\pi d^{2}J_{s}^{d}(1-\sigma _{0}^{2})}{\ell _{B}^{2}%
}\tan ^{2}\Theta
\end{align}%
for $\Theta <\Theta _{\sigma }^{\ast }$. By adjusting the theoretical curve
to the data at the point $\Theta =0$, we fit the data by this curve
throughout the observed range of the tilting angle $\Theta $. As seen in FIG.%
\ref{FigTerasawa}, the fitting is quite good when by assuming constant
values of $N_{\text{spin}}$ and $N_{\text{ppin}}$ throughout the range of $%
\Theta $. A deviation of the theoretical curve from the data for large
tilting angles $\Theta >70^{\circ }$ would be due to effects not taken into
account in the above analysis. For instance, when the parallel magnetic
field become too large, the soliton lattice becomes too dense in the
incommensurate phase and would destabilize skyrmions. We summarize the
numbers $N_{\text{spin}}$ and $N_{\text{ppin}}$ per one
skyrmion-antiskyrmion pair determined by this fitting in a table. As the
sample is tilted, $N_{\text{spin}}$ increases and $N_{\text{ppin}}$
decreases. In particular, both spins and pseudospins are flipped unless $%
\sigma _{0}=0$ or $\sigma _{0}=1$. We conclude that the SU(4) skyrmion
evolves continuously from the ppin-skyrmion limit to the spin-skyrmion limit
as $\sigma _{0}$ increases from $0$ to $1$. 
\begin{equation}
\begin{tabular}{||c|r|r|r|r|r||}
\hline\hline
$\sigma _{0}$ & $0$ & $0.3$ & $0.6$ & $0.7$ & $1$ \\ \hline
$N_{\text{spin}}$ & $0$ & $0.64$ & $2.3$ & $2.8$ & $7.0$ \\ \hline
$N_{\text{ppin}}$ & $1.9$ & $1.8$ & $1.5$ & $0.68$ & $0$ \\ \hline\hline
\end{tabular}
\label{TableSigma}
\end{equation}

\begin{figure}[t]
\begin{center}
\includegraphics[width=0.30\textwidth]{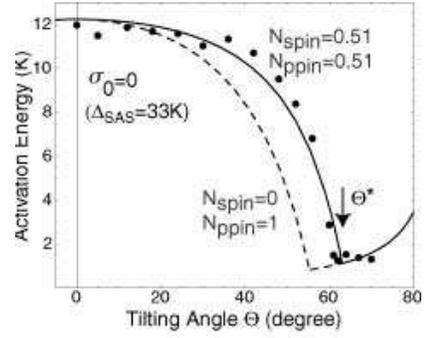}
\end{center}
\caption{ The activation energy at $\protect\nu =1$ is plotted as a function
of $\Theta $ in a sample ($\Delta _{\text{SAS}}=33$K, $\protect\rho %
_{0}=1.0\times 10^{11}$/cm$^{2}$) . The data are taken from Terasawa et al.%
\protect\cite{Terasawa04}. To fit the data by the theoretical formula, we
have adjusted the activation energy at $\Theta =0$ with the experimental
value. We have also assumed that $N_{\text{spin}}$ and $N_{\text{ppin}}$ are
constant for all values of the tilting angle. A better fitting is obtained
as indicated by a solid line when both spins and speudospins are excited.}
\label{FigTera33K}
\end{figure}

It is interesting to study the same problem in a sample having a very large
tunneling gap. Terasawa et al.\cite{Terasawa04} have measured the activation
energy by controlling the tilting angle $\Theta $ at the balanced point in a
sample with $\Delta _{\text{SAS}}\simeq 33$K. We have fitted their data [FIG.%
\ref{FigTera33K}] by the ppin-excitation formula (\ref{TheorMurph}) and by
the generic formula (\ref{TheorTeras}). It is difficult to fit the data if
pure pseudospin excitations are assumed since it is required that $N_{\text{%
ppin}}\geq 1$\ per one pair. A better fitting is obtained if spins and
pseudospins are excited simulaneously since $N_{\text{ppin}}$ may take a
smaller value than 1. Such a simultaneous excitation is allowed at the
balanced point as explained in Appendix \ref{AppenGenerSkyrm}.

In passing we comment on the original mechanism\cite{Moon95B,Yang95B,Read95B}
proposed to explain the activation energy anomaly based on the
exchange-energy loss of bimeron excitations. A bimeron has the same quantum
numbers as a skyrmion, and it can be viewed as a deformed skyrmion with two
meron cores with a string between them. The bimeron excitation energy
consists of the core energy, the string energy and the Coulomb repulsive
energy between the two cores. It is argued that the parallel magnetic field
decreases the string tension and hence the bimeron activation energy.
Clearly the mechanism works well only when the string length is much larger
than the core size. A microscopic calculation has already revealed\cite%
{Brey96B} that the meron core size is large enough to invalidate the naive
picture. Furthermore, the skyrmion is almost as small as the hole itself in
samples with large tunneling gap. On the contrary, in our mechanism the
decrease of the exchange energy follows simply from the phase difference
induced by the parallel magnetic field between the wave functions associated
with the two layers, and it is valid even for small skyrmions.

\section{Discussion}

We have investigated the dynamics of bilayer QH systems based on an
algebraic method inherent to the noncommutative plane with $[X,Y]=-i\ell
_{B}^{2}$. The noncommutativity induced by the LLL projection implies that
the electron position cannot be localized to a point but to a Landau site
occupying an area $2\pi \ell _{B}^{2}$. We have derived the Landau-site
Hamiltonian $H_{\text{C}}$ akin to the lattice Hamiltonian. It has two
entirely different forms, the direct-interaction form $H_{\text{D}}$ and the
exchange-interaction form $H_{\text{X}}$. They are equivalent, $H_{\text{D}%
}=H_{\text{X}}$, as the microscopic Hamiltonian. Nevertheless, the energy of
a charge excitation consists of two well-separated pieces, the direct energy 
$H_{\text{D}}^{\text{cl}}$ and the exchange energy $H_{\text{X}}^{\text{cl}}$%
.

One of our new contributions is the derivation of various LLL-projected
Coulomb potentials in analytic forms. For instance, we have revealed a new
form of the capacitance energy (\ref{CapacFormu}) with $\epsilon _{\text{cap}%
}=4(\epsilon _{\text{D}}^{-}-\epsilon _{\text{X}}^{-})$. It is a sum of the
direct and exchange Coulomb interaction effects, where the direct effect
yields the well-known result from the planar condenser proportional to the
layer separation, while the exchange effect gives a quite large negative
contribution to it when the layer separation is small enough.

We have explored various aspects of SU(4) skyrmions at $\nu =1$. In
particular we have studied a skyrmion-antiskyrmion pair in its small size
limit (electron-hole pair) and its large size limit.

The excitation energy of an electron-hole pair is exactly calculable. We
have obtained the excitation energy as a function of the imbalance parameter 
$\sigma _{0}$. The result is quite interesting: At the balanced point ($%
\sigma _{0}=0$) the pseudospin excitation occurs provided the tunneling gap
is not too large ($\Delta _{\text{SAS}}-\Delta _{\text{Z}}<2\epsilon _{\text{%
X}}^{-}$). A peculiar feature is that, as $\sigma _{0}$ increases, the spin
excitation occurs suddenly because the excitation involves just one electron
or hole. At the monolayer point ($\sigma _{0}=1$) only spins are excited
always. We have also extended the microscopic theory of skyrmions\cite%
{Fertig94B} to our framework. However, a quantitative analysis is yet to be
carried out.

We then have estimated the SU(4) skyrmion excitation energy as a function of 
$\sigma _{0}$ based on the effective Hamiltonian valid for very smooth
isospin textures. In typical samples it flips only pseudospins at the
balanced point. As $\sigma _{0}$ increases, it evolves continuously to flip
both spins and pseudospins, and finally flips only spins at the monolayer
point. We have then calculated how the excitation energy changes as the
sample is tilted. Our formula has explained quite well the activation energy
anomaly found by Murphy et al.\cite{Murphy94L} at the balanced point by
excitations of pseudospins, and also found by Sawada et al.\cite{SawadaX03PE}
at various values of $\sigma _{0}$ by simultaneous excitations of spins and
pseudospins. Though our formulas are derived for sufficiently smooth
skyrmions, they have turned out to be quite good at least with the use of
phenomenological values of $N_{\text{spin}}$ and $N_{\text{ppin}}$. We wish
to develop a microscopic theory to analyze small skyrmions in a future work.
In conclusion, the activation energy anomaly is explained by the loss of the
exchange energy of SU(4) skyrmions, which are reduced always to
spin-skyrmions at the monolayer point and mostly to pseudospin-skyrmions at
the balanced point.

\section*{Acknowledgements}

We would like to thank M. Eliashvili, Y. Hirayama, N. Kumada, D.K.K. Lee, K.
Muraki, A. Sawada and D. Terasawa for fruitful discussions on the subject.
One of the authors (ZFE) is grateful to the hospitality of Theoretical
Physics Laboratory, RIKEN, where a part of this work was done.

\appendix

\section{Group SU(4)}

\label{AppenSU4}

The special unitary group SU(N) has ($N^{2}-1)$ generators. In the standard
representation\cite{BookGellMann}, we denote them as $\lambda _{A}$, $%
A=1,2,\cdots ,N^{2}-1$, and normalize them as%
\begin{equation}
\text{Tr}(\lambda _{A}\lambda _{B})=2\delta _{AB}.  \label{AppNormaL}
\end{equation}%
They are characterized by 
\begin{align}
\lbrack \lambda _{A},\lambda _{B}] =&2if_{ABC}\lambda _{C},  \notag \\
\{\lambda _{A},\lambda _{B}\} =&\frac{4}{N}+\delta _{AB}2d_{ABC}\lambda _{C},
\end{align}%
where $f_{ABC}$ and $d_{ABC}$ are the structure constants of SU(N). We have $%
\lambda _{A}=\tau _{A}$ (the Pauli matrix) with $f_{ABC}=\varepsilon _{ABC}$
and $d_{ABC}=0$ in the case of SU(2).

This standard representation is not convenient for our purpose because the
spin group is SU(2)$\otimes $SU(2) in the bilayer electron system with the
four-component electron field as $\Psi =(\psi ^{\text{f}\uparrow },\psi ^{%
\text{f}\downarrow },\psi ^{\text{b}\uparrow },\psi ^{\text{b}\downarrow })$%
. Embedding SU(2)$\otimes $SU(2) into SU(4) we define the spin matrix by%
\begin{align}
\tau _{x}^{\text{spin}}=& \left( 
\begin{array}{cc}
\tau _{x} & 0 \\ 
0 & \tau _{x}%
\end{array}%
\right) ,\quad \tau _{y}^{\text{spin}}=\left( 
\begin{array}{cc}
\tau _{y} & 0 \\ 
0 & \tau _{y}%
\end{array}%
\right) ,  \notag \\
\tau _{z}^{\text{spin}}=& \left( 
\begin{array}{cc}
\tau _{z} & 0 \\ 
0 & \tau _{z}%
\end{array}%
\right) ,
\end{align}%
and similarly the pseudospin matrix by%
\begin{align}
\tau _{x}^{\text{ppin}}=& \left( 
\begin{array}{cc}
0 & \mathbf{1}_{2} \\ 
\mathbf{1}_{2} & 0%
\end{array}%
\right) ,\quad \tau _{y}^{\text{ppin}}=\left( 
\begin{array}{cc}
0 & -i\mathbf{1}_{2} \\ 
i\mathbf{1}_{2} & 0%
\end{array}%
\right) ,  \notag \\
\tau _{z}^{\text{ppin}}=& \left( 
\begin{array}{cc}
\mathbf{1}_{2} & 0 \\ 
0 & -\mathbf{1}_{2}%
\end{array}%
\right) ,
\end{align}%
where $\mathbf{1}_{2}$\ is the unit matrix in two dimensions. Nine remaining
matrices are products of the spin and pseudospin matrices:%
\begin{align}
\tau _{a}^{\text{spin}}\tau _{x}^{\text{ppin}}=& \left( 
\begin{array}{cc}
0 & \tau _{a} \\ 
\tau _{a} & 0%
\end{array}%
\right) ,\quad \tau _{a}^{\text{spin}}\tau _{y}^{\text{ppin}}=\left( 
\begin{array}{cc}
0 & -i\tau _{a} \\ 
i\tau _{a} & 0%
\end{array}%
\right) ,  \notag \\
\tau _{a}^{\text{spin}}\tau _{z}^{\text{ppin}}=& \left( 
\begin{array}{cc}
\tau _{a} & 0 \\ 
0 & -\tau _{a}%
\end{array}%
\right) .
\end{align}%
Let us denote them as $T_{A}$, $A=1,2,\cdots ,15$, where $T_{1}=T_{x0}$,
etc., $T_{4}=T_{0x}$, etc., $T_{7}=T_{xx}$, $T_{8}=T_{xy}$, etc. with $%
T_{a0}\equiv \tau _{a}^{\text{spin}},\quad T_{0a}\equiv \tau _{a}^{\text{ppin%
}},\quad T_{ab}\equiv \tau _{a}^{\text{spin}}\tau _{b}^{\text{ppin}}$. They
are related with the standard SU(4) generators as%
\begin{align}
T_{x0}& =\lambda _{1}+\lambda _{13},\quad T_{y0}=\lambda _{2}+\lambda _{14},
\notag \\
T_{z0}& =\lambda _{3}-\frac{1}{\sqrt{3}}\lambda _{8}+\frac{\sqrt{6}}{3}%
\lambda _{15},  \notag \\
T_{0x}& =\lambda _{4}+\lambda _{11},\quad T_{0y}=\lambda _{5}+\lambda _{12},
\notag \\
T_{0z}& =\frac{2}{\sqrt{3}}\lambda _{8}+\frac{2}{\sqrt{6}}\lambda _{15}, 
\notag \\
T_{xx}& =\lambda _{6}+\lambda _{9},\quad T_{xy}=\lambda _{10}+\lambda
_{7},\quad T_{xz}=\lambda _{1}-\lambda _{13},  \notag \\
T_{yx}& =\lambda _{10}-\lambda _{7},\quad T_{yy}=\lambda _{6}-\lambda
_{9},\quad T_{yz}=\lambda _{2}-\lambda _{14},  \notag \\
T_{zx}& =\lambda _{4}-\lambda _{11},\quad T_{zy}=\lambda _{5}-\lambda _{12},
\notag \\
T_{zz}& =\lambda _{3}+\frac{1}{\sqrt{3}}\lambda _{8}-\frac{\sqrt{6}}{3}%
\lambda _{15}.  \label{AppBasisSU4}
\end{align}%
The normalization condition reads%
\begin{equation}
\text{Tr}(T_{ab}T_{cd})=4\delta _{ac}\delta _{bd},  \label{AppNormaT}
\end{equation}%
which is different from the standard one (\ref{AppNormaL}).

\section{Decomposition Formula}

\label{AppenDandE}

The Landau-site Hamiltonian $H_{\text{C}}$ possesses two entirely different
forms, the direct-interaction form $H_{\text{D}}$ and the
exchange-interaction form $H_{\text{X}}$. They are equivalent, $H_{\text{D}%
}=H_{\text{X}}$, as the microscopic Hamiltonian. On the other hand, the
excitation energy consists of two well-separated pieces, the direct energy
and the exchange energy. In this appendix we derive the decomposition
formula (\ref{CintoDX}),%
\begin{equation}
\langle \mathfrak{S}|H_{\text{C}}|\mathfrak{S}\rangle =H_{\text{D}}^{\text{cl%
}}+H_{\text{X}}^{\text{cl}},  \label{AppCintoDX}
\end{equation}%
for skyrmion excitations. We also prove the algebraic relation (\ref%
{CasimInX}), or%
\begin{equation}
\sum_{A=1}^{N^{2}-1}\widehat{I}_{A}^{\text{cl}}(\mathbf{x})\bigstar \widehat{%
I}_{A}^{\text{cl}}(\mathbf{x})+\frac{1}{2N}\widehat{\rho }^{\text{cl}}(%
\mathbf{x})\bigstar \widehat{\rho }^{\text{cl}}(\mathbf{x})=\frac{1}{4\pi
\ell _{B}^{2}}\widehat{\rho }^{\text{cl}}(\mathbf{x}),  \label{AppCasimInX}
\end{equation}%
that holds among the classical densities associated with the generators of
the W$_{\infty }$(N) algebra.

We consider a skyrmion state in the SU(N) QH ferromagnet,%
\begin{equation}
|\mathfrak{S}\rangle =\prod_{n=0}\xi ^{\dagger }(n)|0\rangle ,
\label{AppSkyrmState}
\end{equation}%
where%
\begin{equation}
\xi ^{\dagger }(n)=\sum_{\mu }^{N}\left[ u_{\mu }(n)c_{\mu }^{\dagger
}(n)+v_{\mu }(n)c_{\mu }^{\dagger }(n+1)\right] .
\end{equation}%
They satisfy the standard canonical commutation relations,%
\begin{equation}
\lbrack \xi (m),\xi ^{\dagger }(n)]=\delta _{mn},\quad \lbrack \xi (m),\xi
(n)]=0,
\end{equation}%
provided%
\begin{align}
\sum_{\mu }^{N}\left[ u_{\mu }(n)u_{\mu }^{\dagger }(n)+v_{\mu }(n)v_{\mu
}^{\dagger }(n)\right] =& 1,  \notag \\
\sum_{\mu }^{N}v_{\mu }^{\dagger }(n)u_{\mu }(n+1)=& 0.  \label{AppCDXb}
\end{align}%
It follows that%
\begin{equation}
c_{\mu }(n)|\mathfrak{S}\rangle =\left[ u_{\mu }(n)\xi (n)+v_{\mu }(n-1)\xi
(n-1)\right] |\mathfrak{S}\rangle ,
\end{equation}%
from which the only nonvanishing components of two-point correlation
functions are found to be 
\begin{align}
& \langle c_{\mu }^{\ast }(n)c_{\nu }(n)\rangle =u_{\mu }^{\ast }(n)u_{\nu
}(n)+v_{\mu }^{\ast }(n-1)v_{\nu }(n-1),  \notag \\
& \langle c_{\mu }^{\ast }(n)c_{\nu }(n+1)\rangle =u_{\mu }^{\ast }(n)v_{\nu
}(n),  \label{AppCDXc}
\end{align}%
where $\langle c_{\mu }^{\ast }(n)c_{\nu }(n)\rangle =\langle \mathfrak{S}%
|c_{\mu }^{\ast }(n)c_{\nu }(n)|\mathfrak{S}\rangle $\ and so on. We also
derive%
\begin{align}
c_{\mu }(j)c_{\nu }(n)|\mathfrak{S}\rangle =& \left[ u_{\mu }(j)\xi
(j)+v_{\mu }(j-1)\xi (j-1)\right]  \notag \\
& \times \left[ u_{\nu }(n)\xi (n)+v_{\nu }(n-1)\xi (n-1)\right] |\mathfrak{S%
}\rangle .
\end{align}%
Four-point correlation functions $\langle c_{\mu }^{\dag }(m)c_{\sigma
}^{\dag }(i)c_{\tau }(j)c_{\nu }(n)\rangle $ are complicated. Nevertheless,
taking into account the angular-momentum conservation ($V_{mnij}\varpropto
\delta _{m+i,n+j}$) we deduce 
\begin{align}
V_{mnij}& \left\langle c_{\mu }^{\dag }(m)c_{\sigma }^{\dag }(i)c_{\tau
}(j)c_{\nu }(n)\right\rangle  \notag \\
=& V_{mnij}\left\langle c_{\mu }^{\dag }(m)c_{\nu }(n)\right\rangle
\left\langle c_{\sigma }^{\dag }(i)c_{\tau }(j)\right\rangle  \notag \\
& -V_{mnij}\left\langle c_{\mu }^{\dag }(m)c_{\tau }(j)\right\rangle
\left\langle c_{\sigma }^{\dag }(i)c_{\nu }(n)\right\rangle .
\label{AppCDXa}
\end{align}%
This is the general expression allowing to express all kind of Coulomb
energies in the exact form.

It follows from (\ref{AppCDXa}) that%
\begin{align}
& V_{mnij}\sum_{\mu \nu \tau \nu }\left\langle c_{\mu }^{\dag }(m)c_{\sigma
}^{\dag }(i)c_{\tau }(j)c_{\nu }(n)\right\rangle \delta _{\mu \nu }\delta
_{\sigma \tau }  \notag \\
=& V_{mnij}\rho ^{\text{cl}}(m,n)\rho ^{\text{cl}}(i,j)-V_{mnij}X(m,n,i,j),
\end{align}%
where we have set%
\begin{equation}
X(m,n,i,j)\equiv \sum_{\mu \nu \tau \nu }\left\langle c_{\mu }^{\dag
}(m)c_{\tau }(j)\right\rangle \left\langle c_{\sigma }^{\dag }(i)c_{\nu
}(n)\right\rangle \delta _{\mu \nu }\delta _{\sigma \tau }.  \label{Xmnij}
\end{equation}%
Here we use the algebraic identity (\ref{AlgebDX}) to find%
\begin{equation}
X(m,n,i,j)=2\left[ I_{A}^{\text{cl}}(m,j)I_{A}^{\text{cl}}(i,n)+\frac{1}{2N}%
\rho ^{\text{cl}}(m,j)\rho ^{\text{cl}}(i,n)\right] .  \label{AppCDXd}
\end{equation}%
Consequently, 
\begin{align}
E_{\text{C}}=& -\sum_{mnij}{V}_{mnij}\sum_{\sigma ,\tau }\langle c_{\sigma
}^{\dagger }(m)c_{\tau }^{\dagger }(i)c_{\sigma }(n)c_{\tau }(j)\rangle 
\notag \\
=& E_{\text{D}}+E_{\text{X}},
\end{align}%
where%
\begin{align}
E_{\text{D}}=& \sum_{mnij}{V}_{mnij}\rho ^{\text{cl}}(m,n)\rho ^{\text{cl}%
}(i,j), \\
E_{\text{X}}=& -2\sum_{mnij}{V}_{mnij}\big[\sum_{A=1}I_{A}^{\text{cl}%
}(m,j)I_{A}^{\text{cl}}(i,n)  \notag \\
& \hspace{0.8in}+\frac{1}{2N}\rho ^{\text{cl}}(m,j)\rho ^{\text{cl}}(i,n)%
\big].
\end{align}%
Now, $E_{\text{D}}$ and $E_{\text{X}}$ are identical to (\ref{HamilPInD})
and (\ref{SU4InvarMN}) when we replace $\rho (m,n)$ and $I_{A}(m,n)$ with $%
\rho ^{\text{cl}}(m,n)$ and $I_{A}^{\text{cl}}(m,j)$. Hence, we have
established the decomposition formula (\ref{AppCintoDX}) rigorously.

We proceed to prove the algebraic relation (\ref{AppCasimInX}) for the
skyrmion state (\ref{AppSkyrmState}). From (\ref{Xmnij}) we define 
\begin{align}
& X(m,n)\equiv \sum_{ij}X(m,n,i,j)\delta _{ij}  \notag \\
& =\sum_{j}\sum_{\mu \sigma }\left\langle c_{\mu }^{\dag }(m)c_{\sigma
}(j)\right\rangle \left\langle c_{\sigma }^{\dag }(j)c_{\mu
}(n)\right\rangle .
\end{align}%
Using (\ref{AppCDXb}) and (\ref{AppCDXc}) we find the only nonvanishing
values to be given by 
\begin{align}
X(n,n)=& u^{\dag }(n)u(n)+v^{\dag }(n-1)v(n-1),  \notag \\
X(n,n+1)=& [X(n+1,n)]^{\ast }=u^{\dag }(n)v(n),
\end{align}%
which implies that $X(m,n)=\rho ^{\text{cl}}(m,n)$. Combining this with (\ref%
{AppCDXd}) we obtain 
\begin{equation}
I_{A}^{\text{cl}}(m,l)I_{A}^{\text{cl}}(l,n)+\frac{1}{2N}\rho ^{\text{cl}%
}(m,l)\rho ^{\text{cl}}(l,n)=\frac{1}{2}\rho ^{\text{cl}}(m,n).
\end{equation}%
This is equivalent to (\ref{AppCasimInX}) in the coordinate space.

\section{Coulomb Energies}

\label{AppenCouloEnerg}

We estimate the Coulomb self-energy of one skyrmion in the bilayer QH
ferromagnet. It is defined by%
\begin{equation}
E_{\text{self}}=\pi \int \!d^{2}q\,\delta \rho _{\text{sky}}(-\mathbf{q}%
)V^{+}(\mathbf{q})\delta \rho _{\text{sky}}(\mathbf{q})
\end{equation}%
with 
\begin{equation}
V^{+}(q)={\frac{e^{2}}{8\pi \varepsilon q}}\left( 1+e^{-qd}\right) .
\label{AppPotenDP}
\end{equation}%
The density modulation $\delta \rho _{\text{sky}}(\mathbf{x})$ is given by (%
\ref{DensiModul}) for a large skyrmion. Its Fourier component reads%
\begin{equation}
\delta \rho _{\text{sky}}(\mathbf{q})=\frac{1}{2\pi }\int \!d^{2}x\,e^{i%
\mathbf{qx}}\delta \rho _{\text{sky}}(\mathbf{x})=\frac{\alpha q}{2\pi }%
K_{1}(\alpha q),
\end{equation}%
where we have set $\alpha =2\kappa \ell _{B}$. To calculate the energy we
use (\ref{MathFormuA}) and%
\begin{equation}
\int\limits_{0}^{\infty }\!dr\,\frac{r^{n+1}\hspace*{0.25mm}J_{n}(qr)}{%
(r^{2}+\alpha ^{2})^{m+1}}=\frac{q^{m}\alpha ^{n-m}}{2^{m}m!}K_{n-m}(\alpha
q),
\end{equation}%
where $K_{n}(z)=K_{-n}(z)$ is the modified Bessel function: See the formula
(11.4.44) in Ref.\cite{Abramowitz}. Now it is straightforward to derive%
\begin{equation}
E_{\text{self}}=\frac{E_{\text{C}}^{0}}{8\kappa }\int
z^{2}[K_{1}(z)]^{2}\left( 1+e^{-\frac{d}{2\ell }\frac{z}{\kappa }}\right) dz
\label{AppSelfEnerg}
\end{equation}%
as the Coulomb self-energy of one skyrmion.

The SU(4)-noninvariant term yields the capacitance energy (\ref{CapacFormC1}%
), 
\begin{equation}
E_{\text{cap}}={2}\pi \ell _{B}^{2}\epsilon _{\text{cap}}\int
\!d^{2}x\,\delta P_{z}^{\text{sky}}(\mathbf{x})\delta P_{z}^{\text{sky}}(%
\mathbf{x}),  \label{AppCouloE}
\end{equation}%
where $\delta P_{z}^{\text{sky}}(\mathbf{x})$ is given by (\ref{SkyAntiSky}%
). After some calculation we obtain%
\begin{align}
\frac{E_{\text{cap}}}{\epsilon _{\text{cap}}}=& \frac{\sigma _{0}^{2}}{%
24\kappa ^{2}}+(1-\sigma _{0}^{2})\kappa _{p}^{2}A(\kappa )+\sigma
_{0}^{2}\left( \frac{1}{4\kappa ^{2}}-1\right) \frac{\kappa _{p}^{2}}{%
2\kappa ^{2}}  \notag \\
& +\sigma _{0}^{2}\left( \frac{1}{10\kappa ^{2}}-\frac{2}{3}+2\kappa
^{2}\right) \frac{\kappa _{p}^{4}}{\kappa ^{4}},  \label{AppCouloM}
\end{align}%
where $\alpha =2\kappa \ell _{B}$ and $\alpha _{p}=2\kappa _{p}\ell _{B}$: $%
A(\kappa )$ is divergent logarithmically,%
\begin{equation}
A(\kappa )=\frac{1}{\kappa ^{2}}\int_{0}^{\infty }\left( \frac{1}{%
r^{2}+\alpha ^{2}}-\frac{2\alpha ^{2}\ell _{B}^{2}}{\left( r^{2}+\alpha
^{2}\right) ^{3}}\right) ^{2}r^{3}dr.
\end{equation}%
We take the divergent term as the leading contribution for the pseudospin
component,%
\begin{equation}
E_{\text{cap}}\simeq \frac{1}{2}(1-\sigma _{0}^{2})\epsilon _{\text{cap}}N_{%
\text{ppin}}(\kappa _{p})
\end{equation}%
with (\ref{NumbePpinX}).

Finally we present the formula for the exchange energy (\ref{ExchaSU4}) for
a general SU(4) skyrmion (\ref{SU4Skyrm}). After a straightforward but
tedious calculation we obtain%
\begin{align}
& E_{\text{X}}^{\text{SU(4)}}  \notag \\
=& 4\pi J_{s}^{+}\left( 1+\frac{1}{10\kappa ^{4}}\right) -4\pi
J_{s}^{-}\left( \frac{1}{3}+\frac{1-3\sigma _{0}^{2}}{140\kappa ^{4}}\right)
\notag \\
& +4\pi J_{s}^{-}\left( \frac{2}{3}+\frac{3}{14\kappa ^{4}}\right) \left(
\sigma _{0}\frac{\kappa _{s}}{\kappa }+\sqrt{1-\sigma _{0}^{2}}\frac{\kappa
_{r}}{\kappa }\right) ^{2}\frac{\kappa _{s}^{2}}{\kappa ^{2}}  \notag \\
& +4\pi J_{s}^{-}\left( \frac{1+\sigma _{0}^{2}}{3}+\frac{1-19\sigma _{0}^{2}%
}{140\kappa ^{4}}\right) \frac{\kappa _{s}^{2}}{\kappa ^{2}}  \notag \\
& +4\pi J_{s}^{-}(1-\sigma _{0}^{2})\left( \frac{1}{3}+\frac{1}{140\kappa
^{4}}\right) \frac{\kappa _{r}^{2}}{\kappa ^{2}}  \notag \\
& +4\pi J_{s}^{-}\sigma _{0}\sqrt{1-\sigma _{0}^{2}}\left( \frac{2}{3}+\frac{%
9}{70\kappa ^{4}}\right) \frac{\kappa _{s}\kappa _{r}}{\kappa ^{2}}.
\end{align}%
By setting $\kappa _{r}=0$, this is reduced to $E_{\text{X}}^{\text{SU(4)}%
}=E_{\text{X}}+\Delta E_{\text{X}}$ with\ (\ref{IpinExchaEnerg}), where%
\begin{align}
\Delta E_{\text{X}}=& \frac{4\pi }{10\kappa ^{4}}J_{s}^{+}-\frac{4\pi \kappa
_{p}^{2}}{140\kappa ^{6}}J_{s}^{-}  \notag \\
& +\frac{4\pi \sigma _{0}^{2}}{140\kappa ^{8}}\left[ 3\kappa ^{4}-19\kappa
^{2}\kappa _{s}^{2}+30\kappa _{s}^{4}\right] J_{s}^{-}.
\end{align}%
The correction term is small for a large skyrmion.

\section{Continuous Transformation}

\label{AppenGenerSkyrm}

Based on the skyrmion-energy formula (\ref{SkyrmEnergFunct}) we verify that,
if a ppin-skyrmion ($\kappa _{s}=0$, $\kappa _{p}=\kappa )$ is excited at
the balanced point, it evolves continuously into a spin-skyrmion ($\kappa
_{s}=\kappa $, $\kappa _{p}=0)$ via a generic skyrmion ($\kappa _{s}\kappa
_{p}\neq 0)$ as the imbalance parameter $\sigma _{0}$\ increases.

We summarize the total skyrmion energy (\ref{SkyrmEnergFunct}) as a function
of $\kappa =\sqrt{\kappa _{s}^{2}+\kappa _{p}^{2}}$, $z=\left( \kappa
_{s}/\kappa \right) ^{2}$ and $\sigma _{0}$ in the following form, 
\begin{equation}
E_{\text{sky}}(\kappa ,z;\sigma _{0})=\frac{4\pi J_{s}^{+}\sigma _{0}^{2}}{3}%
z^{2}+A(\kappa ;\sigma _{0})z+B(\kappa ;\sigma _{0}),
\end{equation}%
where%
\begin{align}
A(\kappa ;\sigma _{0})=& \frac{4\pi J_{s}^{-}}{3}(1+\sigma _{0}^{2})-\frac{%
1-\sigma _{0}^{2}}{2}\epsilon _{\text{cap}}\kappa ^{2}N_{\xi }  \notag \\
& +\Delta _{\text{Z}}\kappa ^{2}N_{\xi }-\frac{\Delta _{\text{SAS}}}{\sqrt{%
1-\sigma _{0}^{2}}}\kappa ^{2}N_{\xi }
\end{align}%
with (\ref{FuncN}). The explicit expression of $B(\kappa ;\sigma _{0})$ is
not necessary. The variable $z$ is limited in the range $0\leq z\leq 1$.

We start with the balanced point ($\sigma _{0}=0$), where $E_{\text{sky}%
}(\kappa ,z;\sigma _{0})$ is a linear function of $z$. Let us assume%
\begin{equation}
A(\kappa ;\sigma _{0})>0\qquad \text{at\quad }\sigma _{0}=0.
\label{CondiPpinExcit}
\end{equation}%
Then the energy is minimized at $z=0$ or $\kappa _{s}=0$, where
ppin-skyrmions are excited. We now increases $\sigma _{0}$ from $\sigma
_{0}=0$. As far as $A(\kappa ;\sigma _{0})>0$, ppin-skyrmions have the
lowest energy. Due to the $\Delta _{\text{SAS}}$ term, $A(\kappa ;\sigma
_{0})$ decreases and vanishes at $\sigma _{0}=\sigma _{\text{p}}$,%
\begin{equation}
A(\kappa ;\sigma _{0})=0\qquad \text{at\quad }\sigma _{0}=\sigma _{\text{p}}.
\end{equation}%
Then it becomes negative for $\sigma _{0}>\sigma _{\text{p}}$, and the
energy is minimized at $z=z_{\text{min}}$ with%
\begin{equation}
z_{\text{min}}=\frac{3}{8\pi J_{s}^{+}\sigma _{0}^{2}}|A(\kappa ;\sigma
_{0})|.
\end{equation}%
Because $z_{\text{min}}$ increases continuously from $z_{\text{min}}=0$ at $%
\sigma _{0}=\sigma _{\text{p}}$, the spin component ($\kappa _{s}$) is
excited gradually to form a generic skyrmion ($\kappa _{s}\kappa _{p}\neq 0)$%
. The point $z_{\text{min}}$ increases and achieve at $z_{\text{min}}=1$ at $%
\sigma _{0}=\sigma _{\text{s}}$,%
\begin{equation}
|A(\kappa ;\sigma _{0})|=\frac{8\pi J_{s}^{+}\sigma _{0}^{2}}{3}\qquad \text{%
at\quad }\sigma _{0}=\sigma _{\text{s}},  \label{CondiSpinExcit}
\end{equation}%
where the pseudospin component ($\kappa _{p}$) vanishes continuously.

We conclude that, if $A(\kappa ;\sigma _{0})>0$ at $\sigma _{0}=0$,
ppin-skyrmions are excited for $0\leq \sigma _{0}<\sigma _{\text{p}}$,
generic skyrmions are excited for $\sigma _{\text{p}}<\sigma _{0}<\sigma _{%
\text{s}}$, and finally spin-skyrmions are excited for $\sigma _{\text{s}%
}<\sigma _{0}\leq 1$. The transition occurs continuously as illustrated in
FIG.\ref{FigBLSkyTh}, with the critical points $\sigma _{\text{p}}$ and $%
\sigma _{\text{s}}$ being fixed by (\ref{CondiPpinExcit}) and (\ref%
{CondiSpinExcit}). On the other hand, generic skyrmions are excited at the
balanced point if $A(\kappa ;\sigma _{0})<0$ and $z_{\text{min}}<1$ at $%
\sigma _{0}=0$, and spin skyrmions are excited at the balanced point if $%
A(\kappa ;\sigma _{0})<0$ and $z_{\text{min}}\geq 1$ at $\sigma _{0}=0$.



\begin{thebibliography}{99}
\bibitem{NoteCapac} It is found that $\epsilon _{\text{D}}^{-}\simeq 56$K
and $7.6$K while $\epsilon _{\text{X}}^{-}\simeq 23$K and $6.5$K for the
layer separation $d=231$\AA\ and $31$\AA , respectively. (We have taken $%
\rho _{0}=1.2\times 10^{11}$/cm$^{2}$.) In our analysis the capacitance
energy is given by (\ref{CapacFormu}), where $\epsilon _{\text{cap}%
}=4(\epsilon _{\text{D}}^{-}-\epsilon _{\text{X}}^{-})\simeq 132$K and $4.5$%
K for $d=231$\AA\ and $31$\AA , respectively. In some literature\cite%
{Moon95B,Demler99L} it is assumed that $\epsilon _{\text{cap}}=\epsilon _{%
\text{X}}^{-}$. The capacitance energy becomes quite small for a small layer
separation due to the exchange interaction.
\bibitem{NoteSingleSkyrm} In computing the activation energy of a skyrmion
or an antiskyrmion, one needs to know the chemical potential to remove or
add an electron to the system\cite{Moon95B}. However, this problem does not
exist as far as we consider a skyrmion-antiskyrmion pair because the number
of electrons is unchanged. Thus, when we calculate the skyrmion energy, we
actually consider the energy of a skyrmion in a well-separated
skyrmion-antiskyrmion pair.
\end{thebibliography}
\end{document}